\documentclass[11pt,reqno]{amsart}

\usepackage[pdfborder={0 0 0.5 [3 2]}, plainpages=false]{hyperref}%
\usepackage[left=1in,right=1in,top=1in,bottom=1in]{geometry}%
\usepackage{amsmath}
\usepackage{enumerate}
\usepackage{amssymb}                
\usepackage{amsmath}                
\usepackage{amsfonts}
\usepackage{amsthm}
\usepackage{bbm}
\usepackage[table,xcdraw]{xcolor}
\usepackage{booktabs}
\usepackage{svg}
\usepackage{mathtools}
\usepackage{cool}
\usepackage{url}
\usepackage{graphicx,epsfig}
\usepackage{makecell}
\usepackage{array}
\usepackage{verbatim}

\usepackage{subcaption}
\usepackage[capitalize,nameinlink]{cleveref}
\crefname{section}{section}{sections}
\crefname{appendix}{appendix}{appendices}
\crefname{subsection}{subsection}{subsections}
\Crefname{section}{Section}{Sections}
\Crefname{subsection}{Subsection}{Subsections}
\Crefname{appendix}{Appendix}{Appendices}

\Crefname{figure}{Figure}{Figures}

\crefformat{equation}{\textup{#2(#1)#3}}
\crefrangeformat{equation}{\textup{#3(#1)#4--#5(#2)#6}}
\crefmultiformat{equation}{\textup{#2(#1)#3}}{ and \textup{#2(#1)#3}}
{, \textup{#2(#1)#3}}{, and \textup{#2(#1)#3}}
\crefrangemultiformat{equation}{\textup{#3(#1)#4--#5(#2)#6}}%
{ and \textup{#3(#1)#4--#5(#2)#6}}{, \textup{#3(#1)#4--#5(#2)#6}}{, and \textup{#3(#1)#4--#5(#2)#6}}

\Crefformat{equation}{#2Equation~\textup{(#1)}#3}
\Crefrangeformat{equation}{Equations~\textup{#3(#1)#4--#5(#2)#6}}
\Crefmultiformat{equation}{Equations~\textup{#2(#1)#3}}{ and \textup{#2(#1)#3}}
{, \textup{#2(#1)#3}}{, and \textup{#2(#1)#3}}
\Crefrangemultiformat{equation}{Equations~\textup{#3(#1)#4--#5(#2)#6}}%
{ and \textup{#3(#1)#4--#5(#2)#6}}{, \textup{#3(#1)#4--#5(#2)#6}}{, and \textup{#3(#1)#4--#5(#2)#6}}

\crefdefaultlabelformat{#2\textup{#1}#3}

\hypersetup{hidelinks}

\def\R{{\mathbb R}}

\def\C{{\mathbb C}}

\def\calI{{\mathcal I}}
\def\calL{{\mathcal L}}
\def\calH{{\mathcal H}}
\def\Lw{{\mathcal{L}_\omega}}

\DeclareMathOperator{\diag}{diag}
\DeclareMathOperator{\md}{mod}
\DeclareMathOperator{\RR}{\text{Re}}
\DeclareMathOperator{\II}{\text{Im}}

\newtheorem{proposition}{Proposition}

\graphicspath{ {images/} }

\begin{document}

\title{Spatiotemporal dynamics in a twisted, circular waveguide array}

\author{Ross Parker}
\address{Department of Mathematics, Southern Methodist University, 
Dallas, TX 75275, USA}
\email{rhparker@smu.edu}

\author{Yannan Shen} 
\address{Department of Mathematics, University of Kansas, Lawrence, KS 66045, USA}
\email{yshen@ku.edu}

\author{Alejandro Aceves}
\address{Department of Mathematics, Southern Methodist University, 
Dallas, TX 75275, USA}
\email{aaceves@smu.edu}

\author{John Zweck}
\address{Department of Mathematics, The University of Texas at Dallas, 
Richardson, TX 75080, USA}
\email{zweck@utdallas.edu}

\begin{abstract}
We consider the existence and spectral stability of nonlinear discrete localized solutions representing light pulses propagating in a twisted multi-core optical fiber. By considering an even number, $N$, of waveguides, we derive asymptotic expressions for solutions in which the bulk of the light intensity is concentrated as a soliton-like pulses confined to a single waveguide. The leading order terms obtained are in very good agreement with results of numerical computations. Furthermore, as in the model without temporal dispersion, when the twist parameter, $\phi$, is given by $\phi = \pi/N$, these standing waves exhibit optical suppression, in which a single waveguide remains unexcited, to leading order. Spectral computations and numerical evolution experiments suggest that these standing wave solutions are stable for values of the coupling parameter less than a critical value, at which point a spectral instability results from the collision of an internal eigenvalue with the eigenvalues at the origin. This critical value has a maximum when $\phi = \pi/N$.
\end{abstract}

\noindent

\maketitle


\section{Introduction}

Multi-core optical fibers, which are optical waveguides that contain multiple fiber cores within a single cladding, have important applications in diverse fields. In the linear regime, current research targets the possibility of achieving spatial division multiplexing (SDM) at telecommunication wavelengths with the goal of increasing information capacity~\cite{azadeh}. At high intensities, in the nonlinear regime, current research includes generating high power coherent pulses by combining light propagating in each core~\cite{balakin}. As it relates to basic research, this optics platform presents an opportunity to  theoretically~\cite{ace} and experimentally~\cite{minardi} explore localized spatiotemporal dynamics (discrete light bullets) in nonlinear discrete systems, localized soliton-like solutions, and  other structures such as discrete optical vortices~\cite{pryamikov}. While most  research has been on uniform, equally spaced waveguide arrays, recent technological advances which allow for more general configurations have opened new research directions in what is called topological photonics~\cite{ozawa}. This is the case, for example, when a twist is imposed on a circular multi-core fiber, which in particular allows for fine control of diffraction and light transfer in a similar manner to axis bending in linear waveguide arrays~\cite{Longhi2005}. In the linear regime, 
twisted multi-core fibers have applications in distributive sensing, in which position, temperature, strain, and acoustic signals can be detected along the entire length of the fiber~\cite{Gannot2014,Westbrook2017}. Of particular note is shape sensing, in which the shape of the optical fiber can be reconstructed using measurements of light transmission and scattering through the fiber.

\begin{figure}
\begin{center}
\includegraphics[width=5cm]{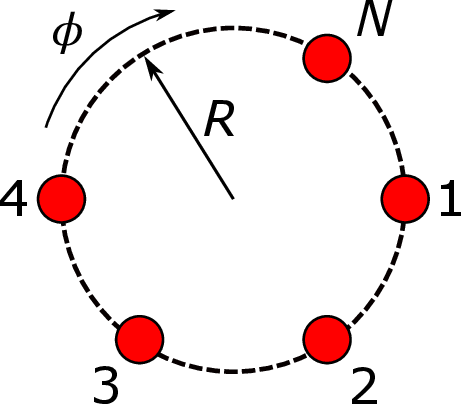}
\end{center}
\caption{Twisted, circular, multi-core fiber consisting of $N$ waveguides.}
\label{fig:circle}
\end{figure}

The coupled mode equations that model optical transmission in twisted, circular, multi-core fibers were derived in \cite{Longhi2007,Longhi2007b,Garanovich2012}. In non-dimensionalized form (and in the setting of no optical gain or loss) these equations are given by
\begin{equation}\label{eq:coupledmode}
i \partial_z c_n + k \left(e^{i\phi}c_{n-1} + e^{-i\phi}c_{n+1} \right) + |c_n|^2 c_n = 0,
\end{equation}
for $n = 1, \dots, N$, where $N$ is the number of fibers in the ring, and the indices $n$ are taken $\md N$ due to the circular geometry (see \cref{fig:circle}).
The complex amplitudes, $c_n(z)$, represent the localized field amplitude in each waveguide, the independent variable, $z$, is along the axis of propagation, and $k$ is the non-dimensional strength of the nearest-neighbor coupling of the waveguides in the ring. The entire fiber structure is twisted along the propagation axis with a spatial period, $\Lambda$, and the twist parameter, $\phi = 4 \pi^2 \epsilon n_s R^2/N \lambda$, is the Peierls phase introduced by the twist \cite{Longhi2007,Peierls1933}, where $\epsilon = 2 \pi / \Lambda$ is the frequency of the twist, $n_s$ the refractive index of the substrate, $R$ is the radius of the circular ring, and $\lambda$ is the wavelength of the propagating field \cite{castro2016,Parto2017}. We note that this model only considers standing wave modes in each waveguide. 
When the twist parameter $\phi = 0$, the model reduces to the discrete nonlinear Schr\"odinger (DNLS) equation \cite{Kevrekidis2009} on a finite, periodic lattice.

Rather than focusing on applications, the results in this paper are aimed at advancing  understanding of the role that topological properties have on the dynamics of spatiotemporal localized modes. 
Indeed, the fiber twist permits the establishment of a topological state in the fiber, in which the intensity in one (or more) of the cores is completely suppressed. Specifically, in prior work, an asymptotic analysis of the optical intensity in a  six-core twisted optical fiber~\cite{castro2016} showed that if the bulk of the optical intensity is confined to a single fiber in the ring, then the intensity in the opposite fiber in the ring is suppressed, to leading order, when $\phi = \pi/6$. This phenomenon can be interpreted as an optical analogue of Aharonov-Bohm (AB) suppression of tunneling in \cite{Ornigotti2007,Parto2017,Parto2019}, in which the fiber twist plays the role of the magnetic flux in the quantum mechanical system \cite{Loss1992}. This suppression effect was established analytically for a four-core optical fiber when $\phi=\pi/4$ in \cite{Parto2019}. In \cite{parker2021} we extended this analysis to standing wave solutions of the form $c_n = a_n e^{i (\omega z + \theta_n) }$ which are constant in amplitude and oscillatory in $z$. We showed that when the number of fibers, $N$, is even, standing wave solutions exist in which the bulk of the intensity is contained in a single fiber core, and the intensity in the opposite core in the ring is completely suppressed for all $z$ when $\phi = \pi/N$. Using numerical computations of spectra and evolution experiments, we showed that these standing wave solutions are stable, and we also extended these results to the case that $N$ is odd.

In this paper, we extend these results to the pulsed regime. To do so, we incorporate a second-order temporal dispersion term, as is found in the nonlinear Schr\"odinger (NLS) equation, namely,
\begin{align}\label{eq:cnz}
&i\partial_z c_n + \partial_t^2 c_n + k\left(e^{i\phi}c_{n-1}+e^{-i\phi}c_{n+1}\right)+|c_n|^2 c_n = 0 && n = 1, \dots, N,
\end{align}
where the complex amplitudes, $c_n$, are now functions of both $z$ and $t$. Using asymptotic analysis, we derive leading order expressions for solutions in which the bulk of the intensity is contained in a single core. Since these results derive from asymptotic series expansions in the coupling parameter, $k$, we consider only the case of weak coupling, i.e. $k < 1$.
We demonstrate that optical suppression also occurs in this model when the number of cores, $N$, is even, and the phase parameter, $\phi$, is given by $\phi=\pi/N$. 

Asymptotic methods have an extensive history as a tool for the analysis of lattice dynamical systems. In a set of pioneering works on the stability of discrete solitons \cite{pelinovsky2005a} and discrete vortices \cite{pelinovsky2005b} in DNLS, the first and second order terms in an asymptotic expansion of the excited sites in the lattice were used to obtain approximations for the eigenvalues of the linearized stability problem (see, for example, \cite[Section 4]{pelinovsky2005a}). While our work builds upon this general method, i.e. we substitute an asymptotic series in the coupling parameter $k$ into the lattice system, and then solve the resulting equations at each order of $k$, there are several key differences. First, the complex amplitudes $c_n$ in \cref{eq:cnz} are functions of $t$ (as well as $z$), thus equation \cref{eq:standingwave} is a set of coupled ordinary differential equations rather than a set of coupled algebraic equations. We are able to simplify these equations using symmetries in the problem, which are a consequence of the circular geometry.
In addition, we use this method to compute the first nonzero term in the asymptotic expansion at every site in the finite lattice, which allows us to obtain a leading order expression for the entire solution in the case where the bulk of the optical intensity is localized to a single site. In particular, since the opposite site in the ring will have the least optical intensity, we can use the asymptotic expansion for this opposite site to demonstrate optical suppression when $\phi=\pi/N$.

The paper is organized as follows. In \cref{sec:model}, we continue our discussion of the mathematical model \cref{eq:cnz} we are studying, %% JZ in particular its Hamiltonian structure, 
and derive equations for pulsed solutions. In \cref{sec:asymp}, we compute the leading order terms in asymptotic expansions of pulsed solutions where the bulk of the intensity is contained in a single waveguide. Details of the computations are deferred to \cref{app:asymp}. In \cref{sec:numerics}, we present numerical results which validate our asymptotic solutions. In \cref{sec:stability}, we investigate the stability of these pulsed solutions, using both spectral computations and evolution simulations, and we end with a brief concluding section which highlights directions for future research.

\section{Mathematical model}\label{sec:model}

We first recast the model \cref{eq:cnz} in vector form. Setting $c = (c_1, \dots, c_n)^T$, we can write \cref{eq:cnz} as
\begin{align}\label{eq:cz}
i \partial_z c + \partial_t^2 c + k A(\phi) c + \diag\left(|c_n|^2\right)c = 0,
\end{align}
where $\diag\left(|c_n|^2\right)$ is the diagonal matrix with diagonal entries $\{|c_1|^2, \dots, |c_N|^2\}$, and $A(\phi)$ is the %% JZperiodic,
tri-diagonal, banded matrix
\begin{align}
A(\phi) = \begin{pmatrix}
0 & e^{-i \phi} & & \dots & e^{i \phi} \\
e^{i \phi} & 0 & e^{-i \phi} & & & \\
& \ddots & \ddots & \ddots &  & \\
 & &e^{i \phi}  & 0 & e^{-i \phi}  \\
e^{-i \phi}& \dots & & e^{i \phi} & 0
\end{pmatrix}.
\end{align}

\begin{comment} %JZ
Equation \cref{eq:cnz} has a Hamiltonian given by
\begin{align}\label{eq:Hc}
\calH(c) = \sum_{n=1}^N \int_{-\infty}^\infty 
\left(
|\dot{c}_n|^2 - \frac{1}{2}|c_n|^4 - e^{i\phi}( c_{n-1}c_n^* + c_{n+1}^* c_n) 
- e^{-i \phi}( c_{n-1}^*c_n + c_{n+1} c_n^*) \right) dt,
\end{align}
where the overdot denotes differentiation with respect to $t$, and the star denotes complex conjugation. While we present our results in this representation, we first point out that one can write the system in standard (real) Hamiltonian form by taking $q_n = \RR c_n$ and $p_n = \II c_n$, so that equation \cref{eq:cnz} becomes
\begin{equation}
\partial_z (q_1, \dots, q_N, p_1, \dots, p_N)^T
 = J \nabla H(q_1, \dots, q_N, p_1, \dots, p_N).
\end{equation}
$J$ is the symplectic matrix
\[
J = \begin{pmatrix}
0 & I_n \\ -I_n & 0
\end{pmatrix},
\]
with $I_n$ the $N \times N$ identity matrix, and where the Hamiltonian \cref{eq:Hc} is written in terms of the $q_n$ and $p_n$ as
\begin{equation}\label{eq:Hqp}
\begin{aligned}
\calH(q_1, \dots, q_N, p_1, \dots, p_N) = \sum_{n=1}^N &\int_{-\infty}^\infty 
\Big(
\dot{q}_n^2 + \dot{p}_n^2 - \frac{1}{2}(q_n^2 + p_n^2)^2 \\
&- \left[(p_{n-1} + p_{n+1}) \cos\phi - (q_{n-1} - p_{n+1}) \sin \phi\right] p_n \\
&- \left[(p_{n-1} - p_{n+1}) \sin\phi + (q_{n-1} + p_{n+1}) \cos \phi\right] q_n
\Big) dt.
\end{aligned}
\end{equation}
\end{comment}

We are interested in finding localized, pulse-like solutions of the form, $c_n(z, t) = C_n(t) e^{i \omega z}$, which are oscillatory in $z$ with wavenumber $\omega$, and have complex amplitude, $C_n(t)$. Substituting this ansatz into \cref{eq:cnz}, the amplitudes $C_n = C_n(t)$ satisfy the system of $N$ coupled ordinary differential equations
\begin{align}\label{eq:standingwave}
\partial_t^2 C_n + k\left(e^{i\phi}C_{n-1}+e^{-i\phi}C_{n+1}\right)+|
C_n|^2 C_n - \omega C_n = 0,
\end{align}
where the subscripts $n$ are taken $\md N$. Letting $C = (C_1, \dots, C_n)^T$, we can write this in matrix form as 
\begin{align}\label{eq:standingwavematrix}
\partial_t^2 C + k A(\phi) C + \diag\left(|C_n|^2 \right)C  - \omega C = 0.
\end{align}
When the coupling parameter is $k=0$, which is known as the anti-continuum (AC) limit \cite{Serg1990,Kevrekidis2009}, equation \cref{eq:standingwavematrix} reduces to $N$ independent copies of the stationary NLS equation, so that the stationary solution at each site is either 0 or the NLS soliton
\begin{equation}\label{eq:NLSsoliton}
\psi(t) = \sqrt{2 \omega} \sech(\sqrt{\omega} t).
\end{equation}
Due to the gauge symmetry of NLS, we can multiply \cref{eq:NLSsoliton} by $e^{i \theta}$ for any $\theta$. 

To properly address the role of the twist, we incorporate a possible nontrivial time dependence to the phase of the field $C_n$ as
\begin{equation}\label{eq:cnansatz}
C_n(t) = a_n(t)e^{i \theta_n(t)},
\end{equation}
where we have separated each complex field $C_n(t)$ into its real amplitude, $a_n(t)$, and phase, $\theta_n(t)$. Substituting \cref{eq:cnansatz} into \cref{eq:standingwave}, we obtain the equation
\begin{align*}
e^{i \theta_n}&\left[ (\ddot a_n - a_n (\dot \theta_n)^2) 
+ i ( a_n \ddot\theta_n + 2 \dot a_n \dot \theta_n ) \right] \\
&+ k\left(e^{i\phi}a_{n-1}e^{i \theta_{n-1}} +e^{-i\phi}a_{n+1}e^{i \theta_{n+1}}\right)+|a_n|^2 a_n e^{i \theta_n} - \omega a_n e^{i \theta_n} = 0,
\end{align*}
where we have suppressed the dependence on $t$. Dividing by $e^{i \theta_n}$, we obtain
\begin{equation}\label{eq:st2}
\begin{aligned}
(\ddot a_n &- a_n (\dot \theta_n)^2) 
+ i ( a_n \ddot\theta_n + 2 \dot a_n \dot \theta_n )\\
&+ k\left(a_{n-1}e^{-i[(\theta_n - \theta_{n-1}) - \phi]} + a_{n+1}e^{i[(\theta_{n+1} - \theta_{n}) - \phi]} \right)+a_n^3 - \omega a_n = 0,
\end{aligned}
\end{equation}  
which we can split up into real and imaginary parts to obtain the system of $2N$ coupled, real-valued ordinary differential equations
\begin{align}
&\ddot a_n - a_n (\dot \theta_n)^2 +
 k\left(a_{n-1}\cos[(\theta_n - \theta_{n-1}) - \phi] + a_{n+1}\cos[(\theta_{n+1} - \theta_{n}) - \phi] \right)+a_n^3 - \omega a_n = 0 \label{eq:st2real} \\
&a_n \ddot\theta_n + 2 \dot a_n \dot \theta_n
+ k\left(-a_{n-1}\sin[(\theta_n - \theta_{n-1}) - \phi] + a_{n+1}\sin [(\theta_{n+1} - \theta_{n}) - \phi] \right) = 0. \label{eq:st2imag} 
\end{align}

\section{Asymptotic analysis}\label{sec:asymp}

\begin{figure}
\begin{center}
\includegraphics[width=7cm]{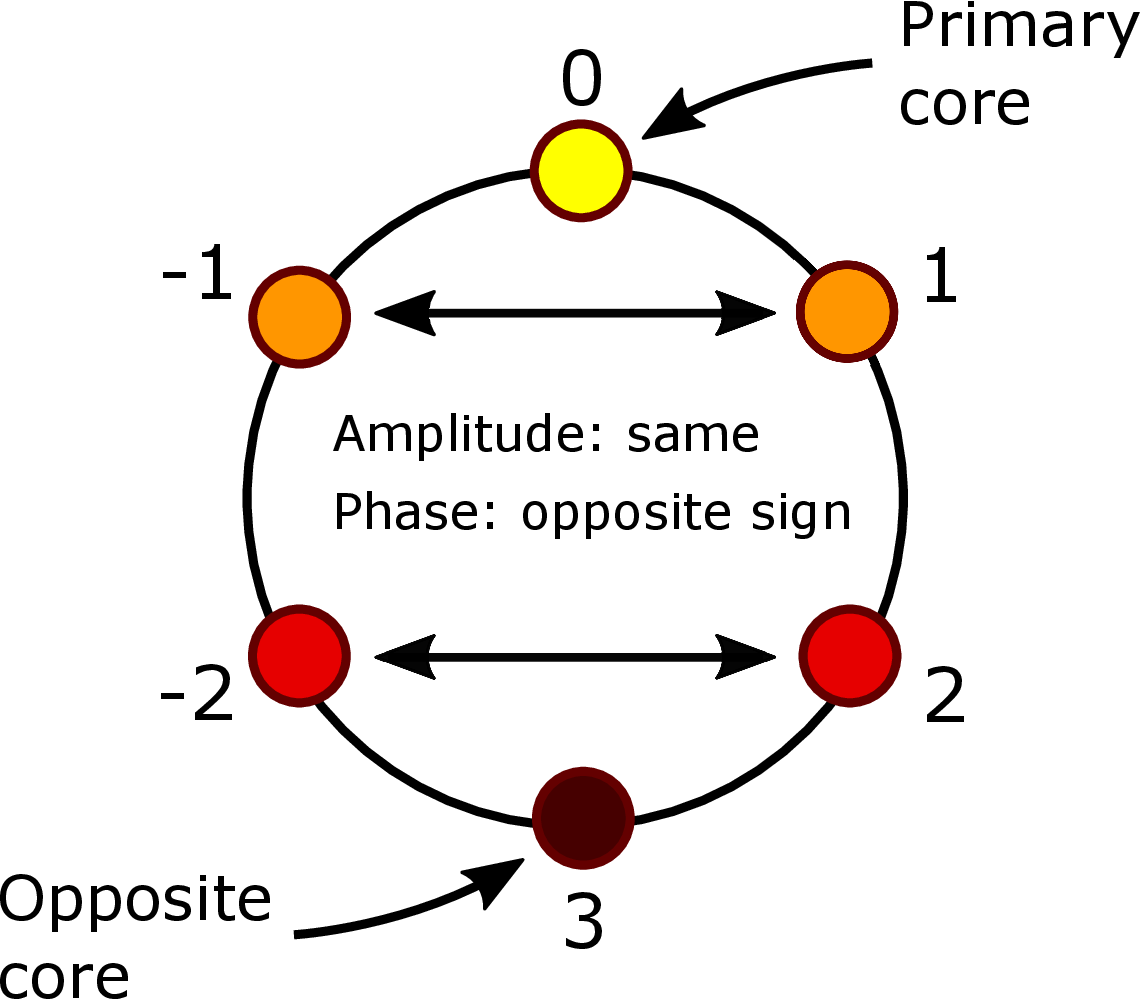}
\end{center}
\caption{Symmetry relations and site labeling in a twisted, circular, multi-core fiber consisting of 6 waveguides.}
\label{fig:symm6}
\end{figure}

We use an asymptotic approach to find highly localized solutions. To do so, we assume that the bulk of the energy is concentrated in a single pulse propagating in one core, which we will call the primary core and label $n=0$. Furthermore, we only consider the case where  the number of waveguides, $N$, is even. We call the waveguide directly opposite the primary core the opposite core, which we label $n=N/2$. The remaining nodes in the ring are labeled $\pm n$, for $n = 1, \dots, N/2-1$, where we count clockwise around the ring for positive $n$ and counterclockwise for negative $n$. See \cref{fig:symm6} for the labeling scheme for $N=6$. Numerical parameter continuation experiments suggest that for $N$ even, there exist solutions with the following symmetries (see \cref{fig:symm6} for an illustration when $N=6$):
\begin{equation}\label{eq:symm}
\begin{aligned}
a_{-n}(t) &= a_{n}(t) && \qquad n = 1, \dots, N/2 \\
\theta_{-n}(t) &= -\theta_{n}(t) && \qquad n = 1, \dots, N/2 \\
\theta_0(t) &= 0 \\
\theta_{N/2}(t) &= 0.
\end{aligned}
\end{equation} 
We verify in \cref{app:symm} that these symmetry conditions are consistent with \cref{eq:st2}, and we will look for a solution with these specific symmetries. 

To find a leading order approximation for a solution to \cref{eq:st2}, we assume that the amplitudes $a_n(t)$ and phases $\theta_n(t)$ have power series expansions in the coupling parameter $k$, where we take $0 \leq k < 1$. Furthermore, we assume that these terms have the following orders of magnitude in $k$, which are suggested by numerical parameter continuation experiments:
\begin{equation}\label{eq:basicseries}
\begin{aligned}
a_0(t) &= \psi(t) + \mathcal{O}\left(k\right) \\
a_n(t) &= \mathcal{O}\left(k^n\right) && n = 1, \dots, N/2 \\
\theta_n(t) &= n \phi + \mathcal{O}\left(k^{N - 2n}\right) && n = 1, \dots, N/2-1.
\end{aligned}
\end{equation}
Since we wish node 0 to contain the bulk of the intensity, the leading order term for $a_0(t)$ is the NLS soliton $\psi(t)$, and the rest of the terms are of order $k$ or higher. The form of equation \cref{eq:st2} suggests that $a_n = \mathcal{O}(k a_{n-1})$, from which we obtain the ansatz for $a_n$ in \cref{eq:basicseries}. It follows from the ansatz for $\theta_n$ that, for $n = 1, \dots, N/2-1$, we have $\theta_n - \theta_{n-1} - \phi = \mathcal{O}\left(k^{N - 2n}\right)$. This implies that, to leading order, the cosine terms in \cref{eq:st2real} and the sine terms in \cref{eq:st2imag} are 1 and 0, respectively.

\subsection{Asymptotic formulae}

Following the procedure detailed in \cref{app:asymp}, we obtain a solution to \cref{eq:st2} of the form
\begin{equation}\label{eq:asympsol}
\begin{aligned}
a_0 &= \psi + k^2 \widetilde{a}_0 + \mathcal{O}(k^3) \\
a_n &= k^n \widetilde{a}_n + \mathcal{O}(k^{n+2}) && n = 1, \dots, N/2 \\
\theta_n &= n \phi + k^{N - 2n} \widetilde{\theta}_n
+ \mathcal{O}(k^{N - 2n+1}) && n = 1, \dots, N/2-1,
\end{aligned}
\end{equation}
where $\widetilde{a}_n$ is defined recursively as 
\begin{equation}\label{eq:tildean}
\begin{aligned}
\widetilde{a}_1 &= (\omega - \partial_t^2)^{-1} \psi \\
\widetilde{a}_n &= (\omega - \partial_t^2)^{-1} \widetilde{a}_{n-1} && n = 2, \dots, N/2-1 \\
\widetilde{a}_{N/2} &= 2 \cos( N\phi/2)(\omega - \partial_t^2)^{-1} \widetilde{a}_{N/2-1}.
\end{aligned}
\end{equation}
The remaining term, $\widetilde{a}_0$, from node 0 can be found by solving the equation
\begin{equation}\label{eq:tildea0eq}
\left( \partial_t^2 - \omega + 3 \psi^2 \right) \widetilde{a}_0 = -2 \widetilde{a}_1 =
-2 (\omega - \partial_t^2)^{-1} \psi,
\end{equation}
with the condition that $\widetilde{a}_0 \perp \dot \psi$ in $L^2(\R)$.
We observe that each function $\widetilde{a}_n$ is defined in terms of $\widetilde{a}_{n-1}$, i.e. the amplitude of each waveguide is defined in terms of the waveguide which is one closer to the primary core.
The invertibility of the linear operator $(\omega - \partial_t^2)$ is discussed in \cref{app:asymp}. Collapsing these recursion relations, we can write \cref{eq:tildean} in terms of the NLS soliton $\psi$ as
\begin{equation}\label{eq:tildeanpsi}
\begin{aligned}
\widetilde{a}_n &= (\omega - \partial_t^2)^{-n} \psi && n = 1, \dots, N/2-1 \\
\widetilde{a}_{N/2} &= 2 \cos( N\phi/2)(\omega - \partial_t^2)^{-N/2} \psi.
\end{aligned}
\end{equation}
Note that $\widetilde{a}_n$ does not depend on $\phi$, except when $n=N/2$. 

For the phases, we obtain the following recursive formulas for the products $\widetilde{a}_n\widetilde{\theta}$:
\begin{equation}\label{eq:tildeantn}
\begin{aligned}
\widetilde{a}_{N/2-1}\widetilde{\theta}_{N/2-1} &= -\sin(N \phi/2) (\omega - \partial_t^2)^{-1} \widetilde{a}_{N/2} \\
\widetilde{a}_n \widetilde{\theta}_n &= (\omega - \partial_t^2)^{-1} \left( \widetilde{a}_{n+1} \widetilde{\theta}_{n+1} \right) && n = 1, \dots, N/2-2,
\end{aligned}
\end{equation}
where each product $\widetilde{a}_n \widetilde{\theta}_n$ is defined in terms of the product $\widetilde{a}_{n+1} \widetilde{\theta}_{n+1}$, i.e. the phase of each waveguide is defined in terms of the waveguide which is one further from the primary core.
Note that these involve the terms $\widetilde{a}_n$, which were computed above. For $n = 1, \dots, N/2-2$, we can again collapse these recurrence relations and write the expression for $\widetilde{a}_n \widetilde{\theta}_n$ in terms of $\widetilde{a}_{N/2-1} \widetilde{\theta}_{N/2-1}$ as
\begin{align*}
\widetilde{a}_n \widetilde{\theta}_n &= (\omega - \partial_t^2)^{-(N/2-n-1)} \left( \widetilde{a}_{N/2-1} \widetilde{\theta}_{N/2-1} \right) && n = 1, \dots, N/2-2.
\end{align*}
Substituting \cref{eq:tildeanpsi} for $\widetilde{a}_n$ and simplifying, we can again write everything in terms of $\psi$ to obtain
\begin{equation}\label{eq:tildeantnpsi}
\begin{aligned}
\widetilde{a}_n \widetilde{\theta}_n &= -\sin(N \phi) (\omega - \partial_t^2)^{-(N-n)} \psi && n = 1, \dots, N/2-1.
\end{aligned}
\end{equation}

\subsection{Optical suppression}

The results of the previous section hold for all cores in the fiber. We now present more refined results for the opposite core.
When $\phi = \pi/N$, it follows from \cref{eq:tildeanpsi} that $\widetilde{a}_{N/2}(t) = 0$ for all $t$. Thus, as in the case without temporal dispersion, the intensity at the opposite core $n=N/2$ in the ring is suppressed at this value of the twist parameter $\phi$. 
It is important to note, however, that this choice of $\phi$ only zeros out the leading order term in the asymptotic expansion \cref{eq:asympsol} for $a_{N/2}$, i.e. the term of $\mathcal{O}(k^{N/2})$, which leaves us with $a_{N/2} = \mathcal{O}(k^{N/2+2})$. By examining higher order terms (see \cref{sec:hot}), we can obtain the stronger result that all terms in the asymptotic expansion are zero up to $\mathcal{O}(k^{N/2+3})$, leaving us with $a_{N/2} = \mathcal{O}(k^{N/2+4})$. The intensity at the opposite core may not be completely suppressed due to the presence of these higher order terms in the asymptotic expansion for $a_{N/2}$ that are much more difficult to compute.
Similarly, it follows from \cref{eq:tildeantnpsi} that $\widetilde{a}_n \widetilde{\theta}_n = 0$ for all $n$ when $\phi = \pi/N$, which zeros out the leading order term in the expansions for $\theta_n$ for all $n$. These findings are confirmed numerically in the next section. We will see that numerical evidence in fact supports complete suppression of the opposite core when $\phi = \pi/N$.

\section{Numerical results}\label{sec:numerics}

\begin{figure}
    \centering
    \includegraphics[width=16cm]{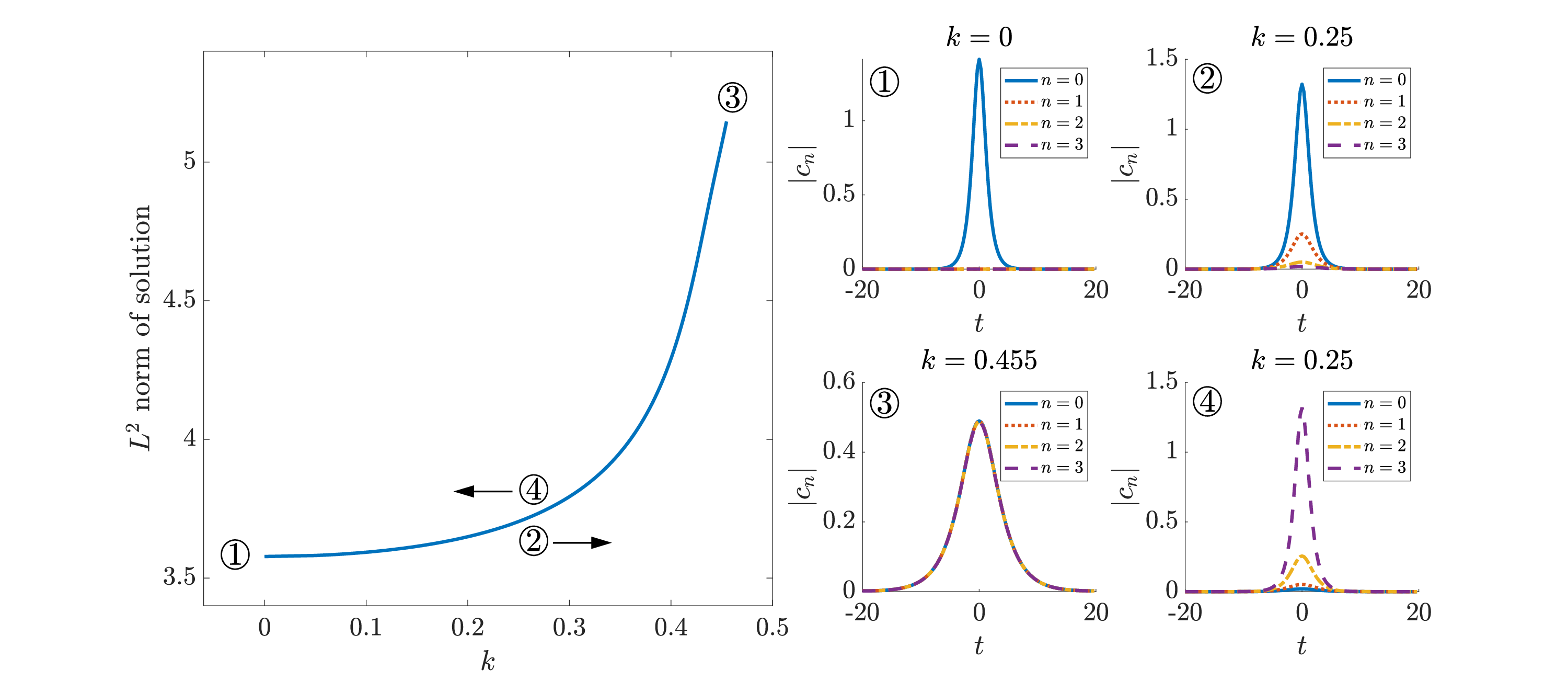}
    \caption{Parameter continuation in $k$ for solutions to \cref{eq:standingwave} for $\phi=0.25$. Left panel: Continuation diagram plotting $L^2$ norm of entire solution vs. $k$. Right panels: Amplitudes at the first four sites of representative solutions, which correspond to labeled points on the continuation diagram. Label 2 is at $k=0.25$ in the direction of increasing $k$. Label 4 is at $k=0.25$ in the direction of decreasing $k$. 128 Fourier nodes per core.}
    \label{fig:kcont}
\end{figure}

\begin{figure}
    \centering
    \begin{subfigure}{0.3\linewidth}
        \caption{}
        \label{fig:kcont2a}
        \includegraphics[width=5cm]{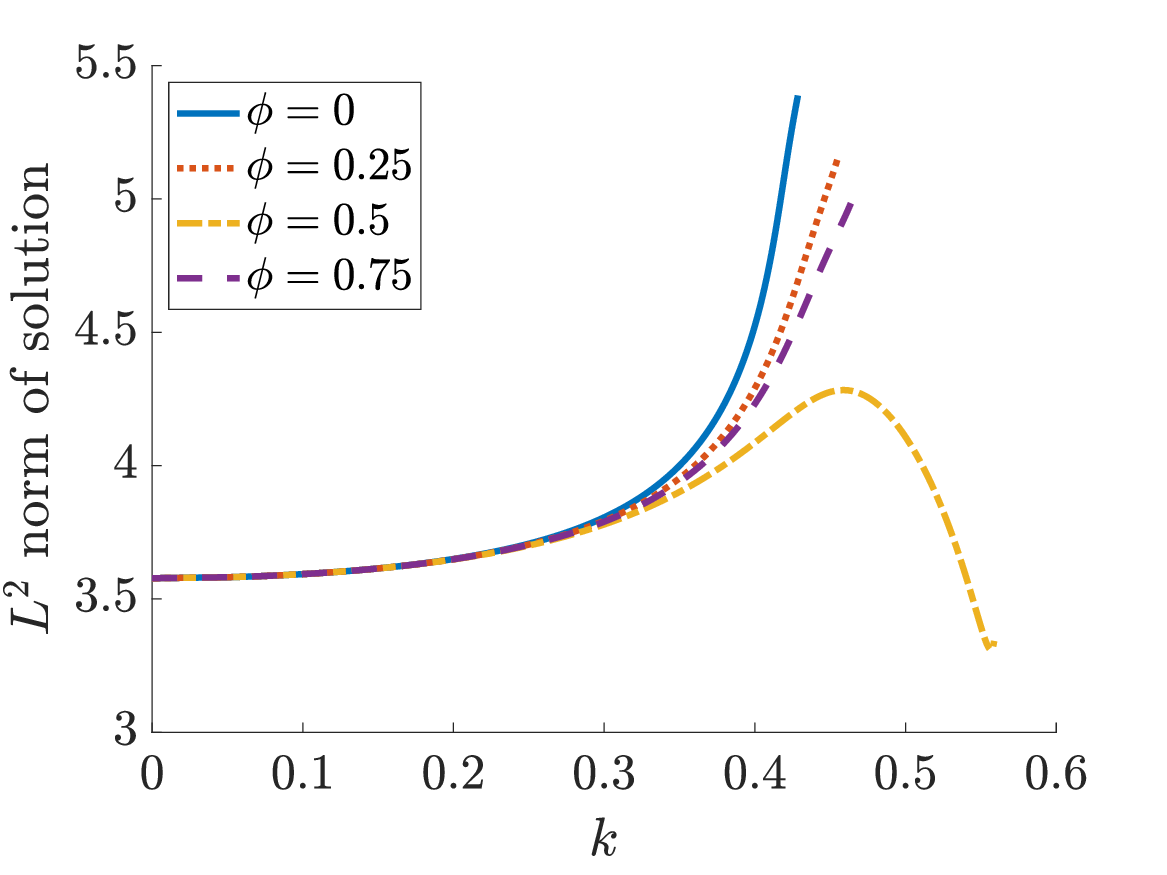}
    \end{subfigure}
    \begin{subfigure}{0.3\linewidth}
        \caption{}
        \label{fig:kcont2b}
        \includegraphics[width=5cm]{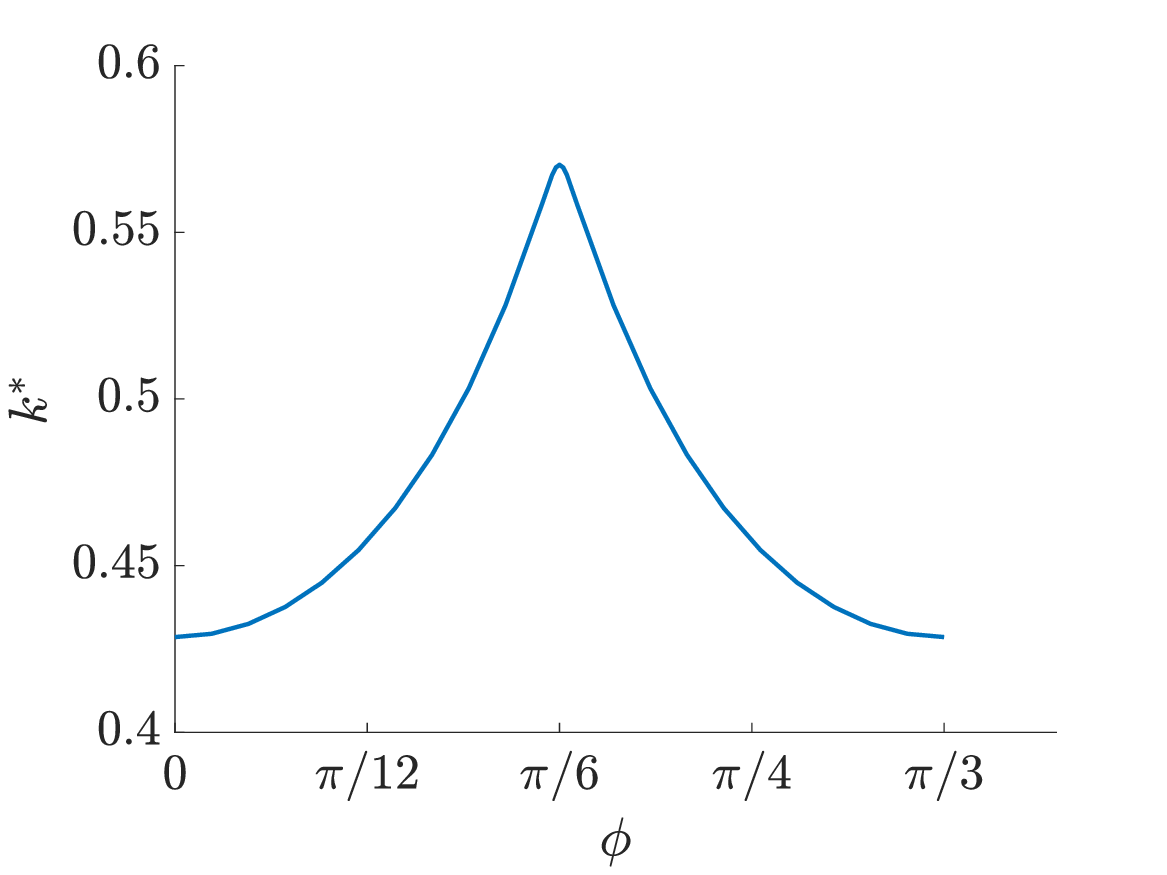}
    \end{subfigure}
    \begin{subfigure}{0.3\linewidth}
        \caption{}
        \label{fig:kcont2c}
        \includegraphics[width=5cm]{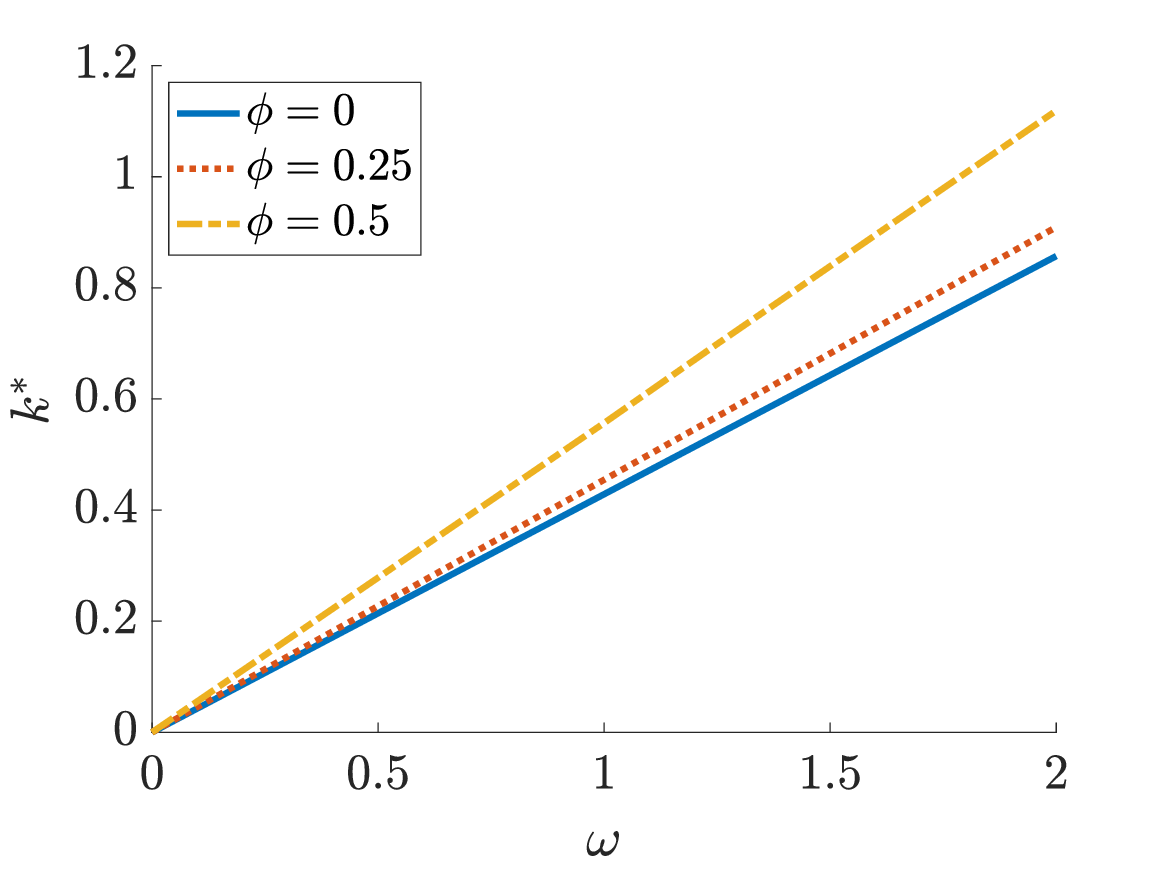}
    \end{subfigure}
    \caption{Parameter continuation in $k$ for solutions to \cref{eq:standingwave}. (a) Continuation diagram plotting $L^2$ norm of entire solution vs. $k$ for various $\phi$. (b) Plot of critical value $k^*$ vs. $\phi$ for $\omega = 1$. (c) Plot of critical value $k^*$ vs $\omega$ for various $\phi$. 128 Fourier nodes per core.}
    \label{fig:kcont2}
\end{figure}

\begin{figure}
    \centering
    \begin{subfigure}{0.3\linewidth}
        \caption{}
        \label{fig:m6suppa3}
        \includegraphics[width=5cm]{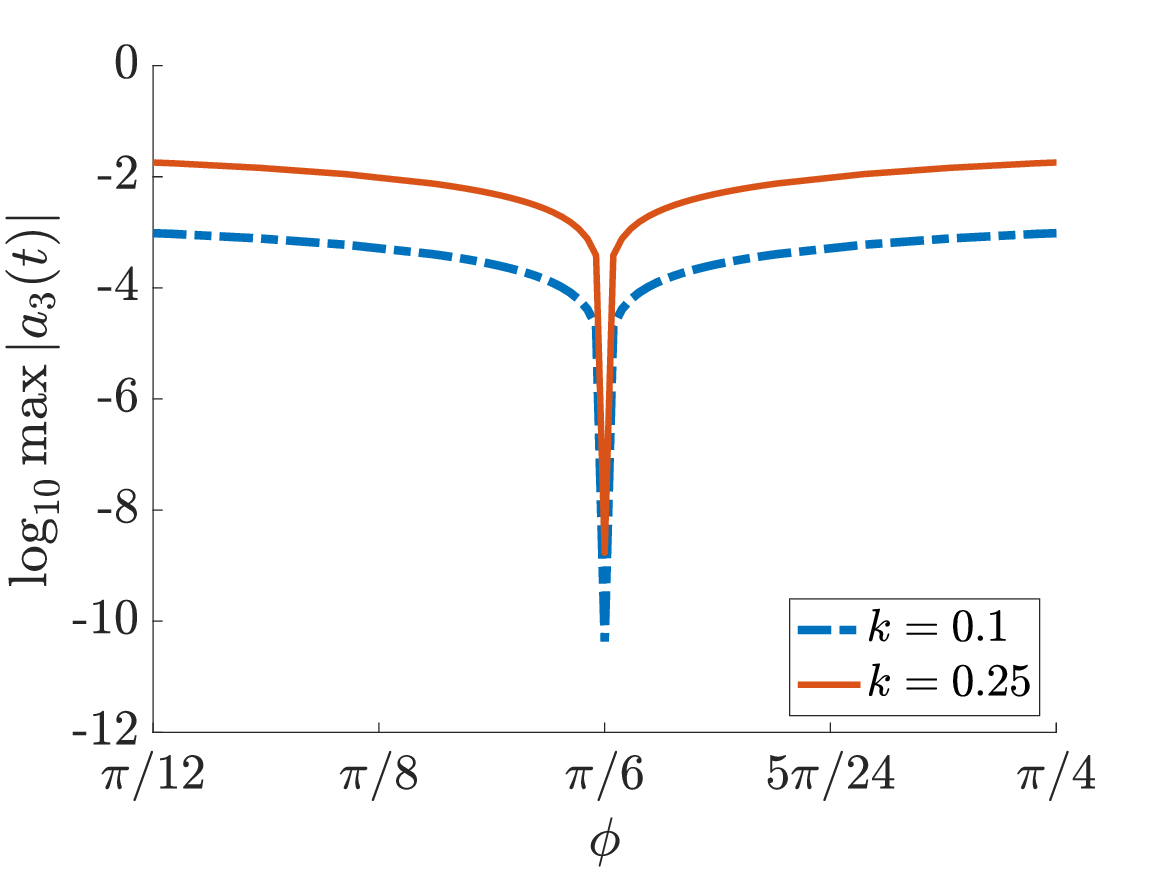}
    \end{subfigure}
    \begin{subfigure}{0.3\linewidth}
        \caption{}
        \label{fig:m6supptheta}
        \includegraphics[width=5cm]{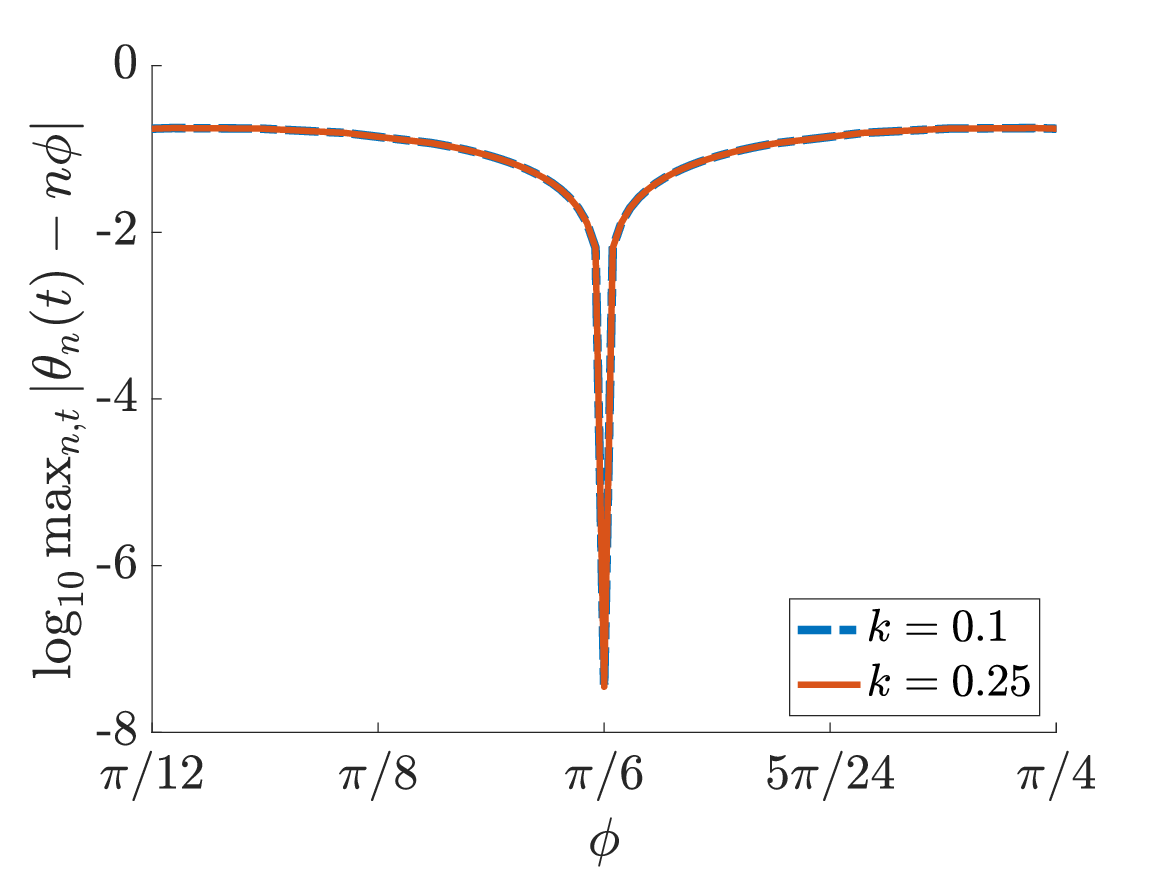}
    \end{subfigure}
    \begin{subfigure}{0.3\linewidth}
        \caption{}
        \label{fig:suppN}
        \includegraphics[width=5cm]{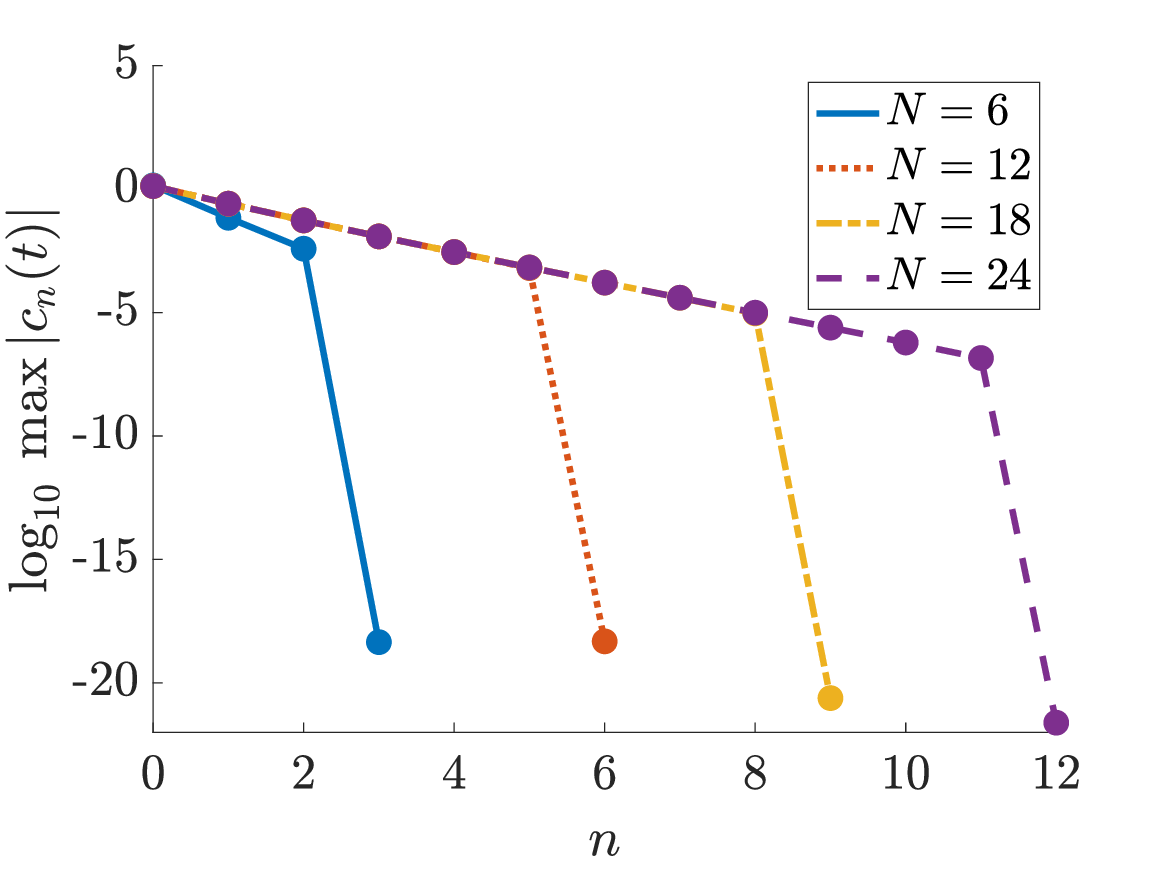}
    \end{subfigure}
    \caption{(a) $\log_{10} \max |a_3(t)|$ vs. $\phi$ for $k=0.1$ and $k=0.25$. (b) $\log_{10} \max_{n,t}{|\theta_n(t) - n \phi|}$ vs. $\phi$ for $k=0.1$ and $k=0.25$. We note that this maximum is taken over all nodes $n$.
    (c) $\log_{10} \max |c_n(t)|$ vs. lattice site $n$ for varying $N$, with twist parameter $\phi=\pi/N$, $k=0.25$, and 128 Fourier nodes.}
    \label{fig:m6supp}
\end{figure}

For all simulations, unless otherwise indicated, we will take $N=6$ fibers and $\omega=1$. 
The choice of $N=6$ fibers is for efficiency of computation, and to parallel what was done in \cite{castro2016,parker2021}. Results for larger $N$ will be shown to illustrate the suppression effect in the opposite core when the twist parameter $\phi$ is $\pi/N$ (\cref{fig:suppN}).
First, we construct pulse solutions to \cref{eq:standingwave} by using parameter continuation from the anti-continuum (AC) limit ($k=0$). To do this, we take $C_n = u_n + i v_n$ in \cref{eq:standingwave}, and separate real and imaginary parts to obtain
\begin{equation}\label{eq:standingwavesep}
\begin{aligned}
&\partial_t^2 u_n + k\left( (u_{n-1} + u_{n+1}) \cos \phi  - (v_{n-1} - v_{n+1})\sin \phi \right) + (u_n^2+v_n^2) u_n - \omega u_n= 0, \\
&\partial_t^2 v_n + k\left( (v_{n-1} + v_{n+1} ) \cos \phi + (u_{n-1}- u_{n+1})\sin \phi \right) +(u_n^2+v_n^2) v_n - \omega v_n = 0.
\end{aligned}
\end{equation}
Letting $u = (u_1, \dots, u_N)^T$ and $v = (v_1, \dots, v_N)^T$, we can write \cref{eq:standingwavesep} in matrix form as 
\begin{equation}\label{eq:standingwavematrixsep}
\begin{aligned}
&\partial_t^2 u + k (A_c(\phi) u - A_s(\phi) v) + \diag\left(u_n^2 + v_n^2 \right)u - \omega u = 0, \\
&\partial_t^2 v + k (A_c(\phi) v + A_s(\phi) u) + \diag\left(u_n^2 + v_n^2 \right)v - \omega v = 0,
\end{aligned}
\end{equation}
where
\begin{align}
A_c(\phi) &= \RR A(\phi) = \begin{pmatrix}
0 & \cos \phi & & \dots & \cos \phi \\
\cos \phi & 0 & \cos \phi & & & \\
& \ddots & \ddots & \ddots &  & \\
 & &\cos \phi  & 0 & \cos \phi  \\
\cos \phi& \dots & & \cos \phi & 0
\end{pmatrix} \label{eq:realA} \\ 
A_s(\phi) &= \II A(\phi) = \begin{pmatrix}
0 & -\sin \phi & & \dots & \sin \phi \\
\sin \phi & 0 & -\sin \phi & & & \\
& \ddots & \ddots & \ddots &  & \\
 & &\sin \phi  & 0 & -\sin \phi  \\
-\sin \phi& \dots & & \sin \phi & 0
\end{pmatrix}.\label{eq:imagA}
\end{align}
To start the continuation, we take $u_0(t) = \psi(t)$, $u_n(t) = 0$ for $n \neq 0$, and $v_n(t) = 0$ for all $n$, i.e. we start with a real NLS soliton in node 0, and the zero solution everywhere else. Spatial discretization is done using a Fourier spectral discretization with periodic boundary conditions on the interval $[-T,T]$, where $T$ is chosen sufficiently large so that the tails of the localized solutions have sufficient room to decay. Unless otherwise indicated, we will take $T=20$, and use 256 Fourier nodes per fiber core. Results of the parameter continuation for $\phi = 0.25$ are shown in \cref{fig:kcont}. As $k$ increases, the intensity spreads from the primary core at $n=0$ to the other cores in the ring (label 2 in \cref{fig:kcont}). A turning point is reached at a critical value $k^*$ (approximately 0.455 in \cref{fig:kcont}; see \cref{fig:kcont2a} for other values of $\phi$), at which point 
$k$ has reached a maximum, and
all cores have equal amplitudes (label 3 in \cref{fig:kcont}). The parameter continuation then reverses direction. Solutions for decreasing $k$ are the same as those for increasing $k$, except that the primary core is now located at $n=N/2$ (compare labels 2 and 4 in \cref{fig:kcont}; the amplitudes are identical but are located at different sites in the ring). This critical value $k^*$ depends on both $\phi$ and $\omega$. Numerical continuation experiments suggest that, for fixed $\omega$, the maximum value of $k^*$ occurs when $\phi = \pi/N$ (see \cref{fig:kcont2b} for results for $N=6$). Numerical continuation experiments also suggest that, for fixed $\phi$, $k^*$ is proportional to $\omega$ (\cref{fig:kcont2c}). Parameter continuation results for larger $N$ (up to $N=24$) are similar; however, since the discretization requires $NM$ points, where $M$ is the number of Fourier nodes per waveguide, this is computationally infeasible for very large $N$.

\begin{figure}
    \centering
    \begin{subfigure}{0.3\linewidth}
        \caption{}
        \label{fig:m6025amp}
        \includegraphics[width=5cm]{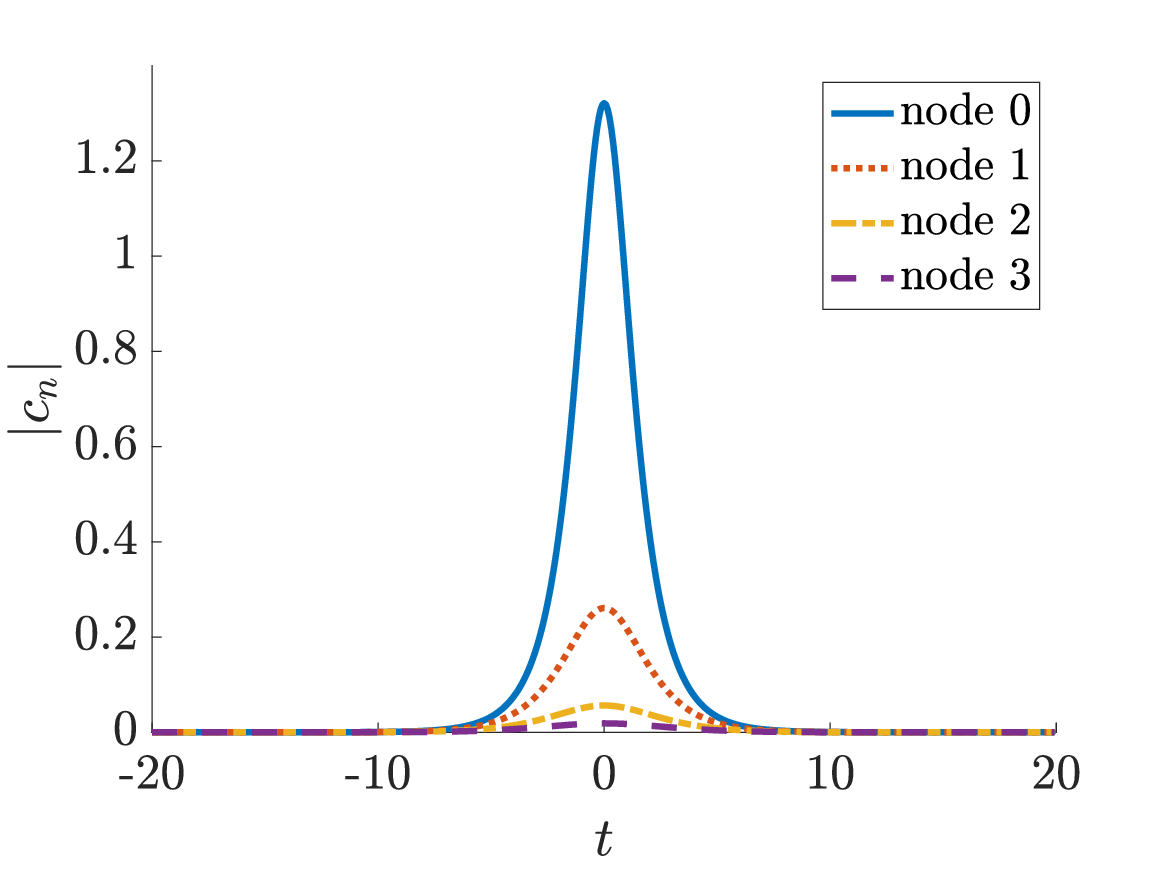}
    \end{subfigure}
    \begin{subfigure}{0.3\linewidth}
        \caption{}
        \label{fig:m6025logamp}
        \includegraphics[width=5cm]{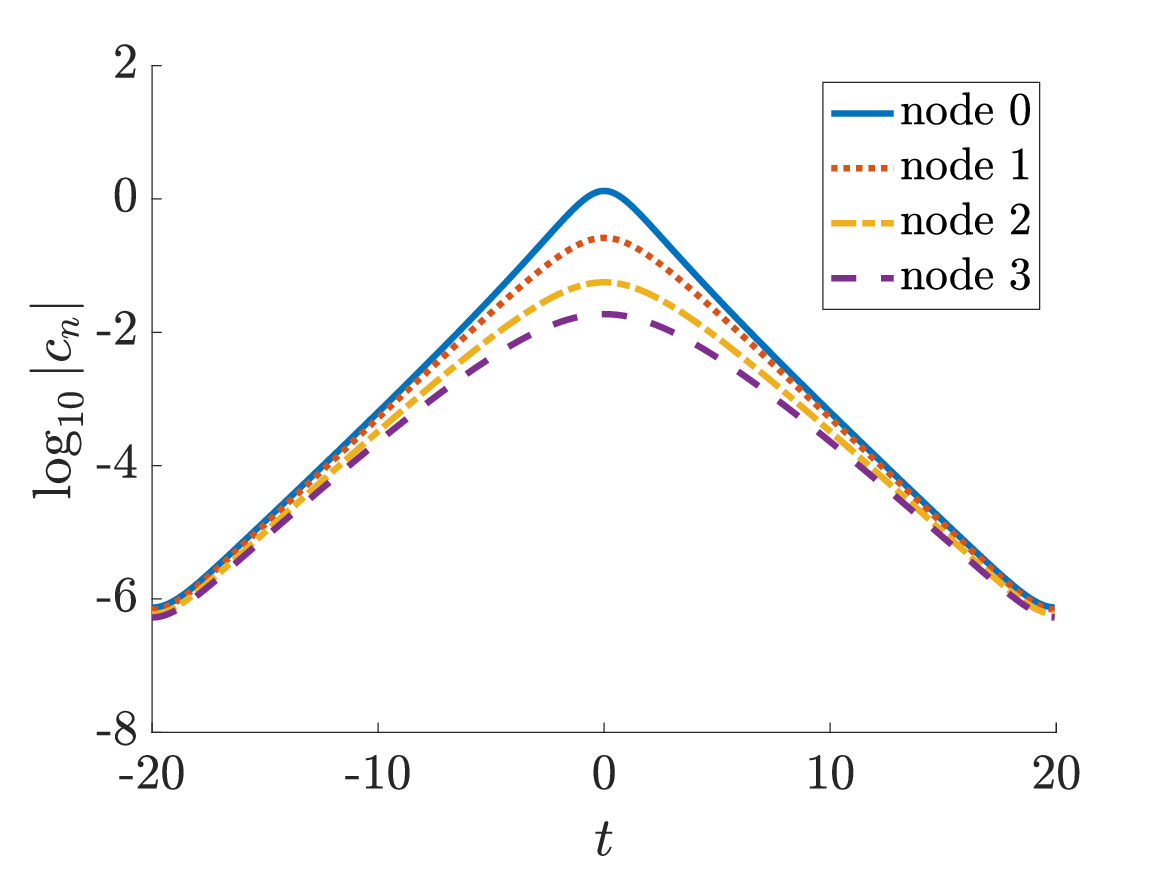}
    \end{subfigure}
    \begin{subfigure}{0.3\linewidth}
        \caption{}
        \label{fig:m6025phase}
        \includegraphics[width=5cm]{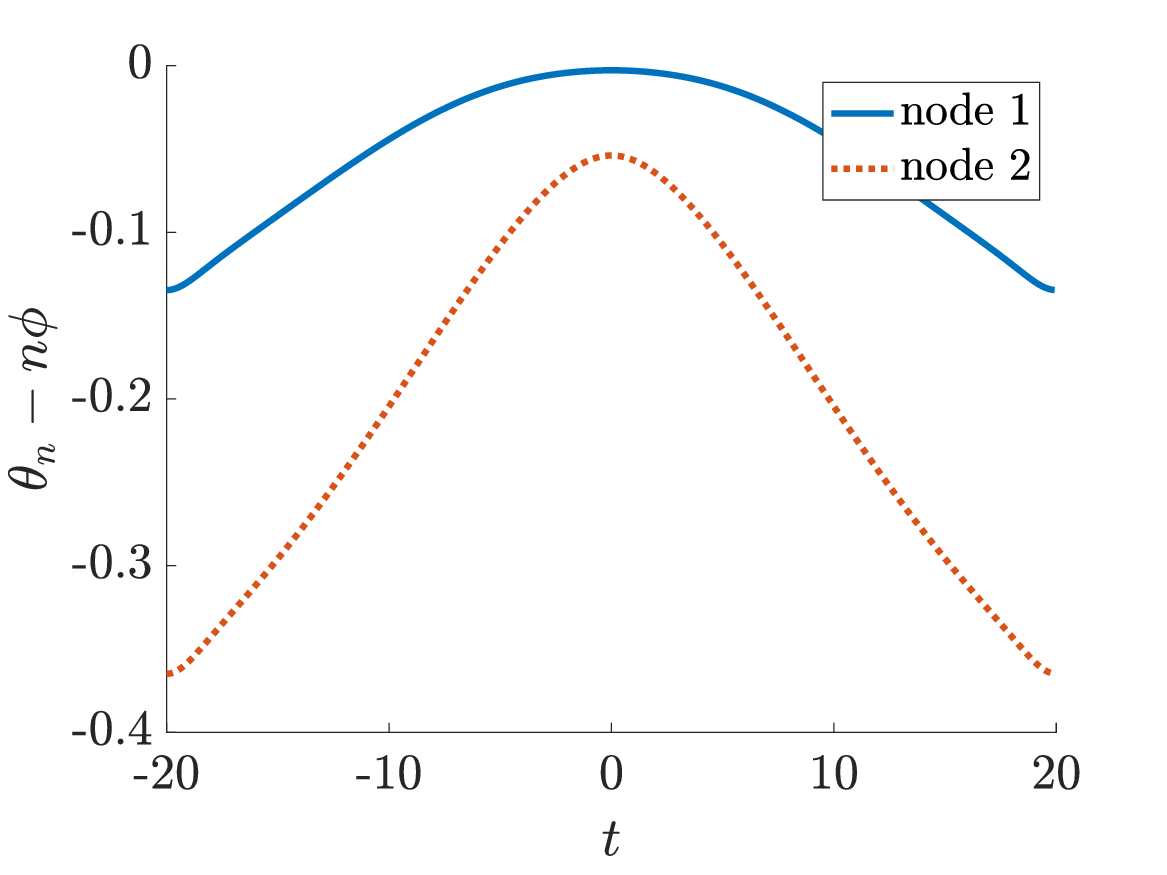}
    \end{subfigure}
    \begin{subfigure}{0.3\linewidth}
        \caption{}
        \label{fig:m6pi6amp}
        \includegraphics[width=5cm]{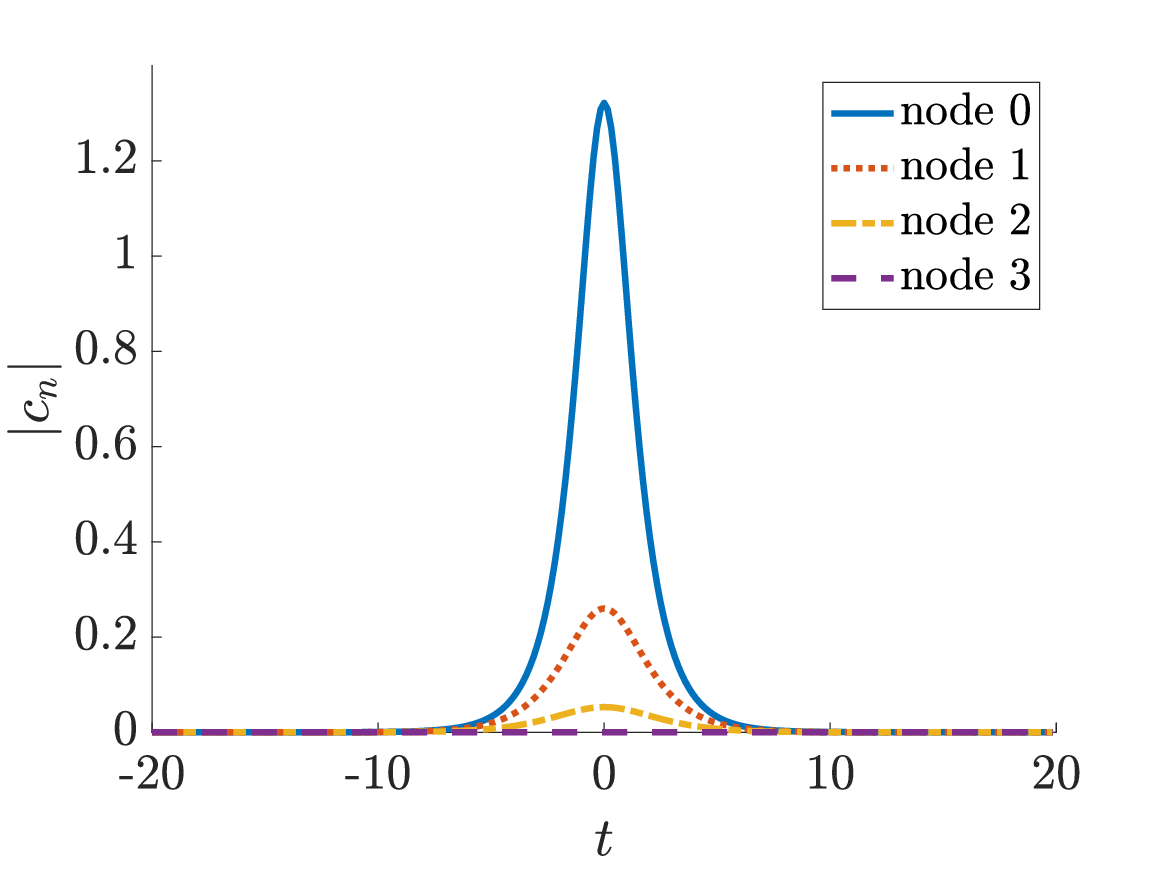}
    \end{subfigure}
    \begin{subfigure}{0.3\linewidth}
        \caption{}
        \label{fig:m6pi6logamp}
        \includegraphics[width=5cm]{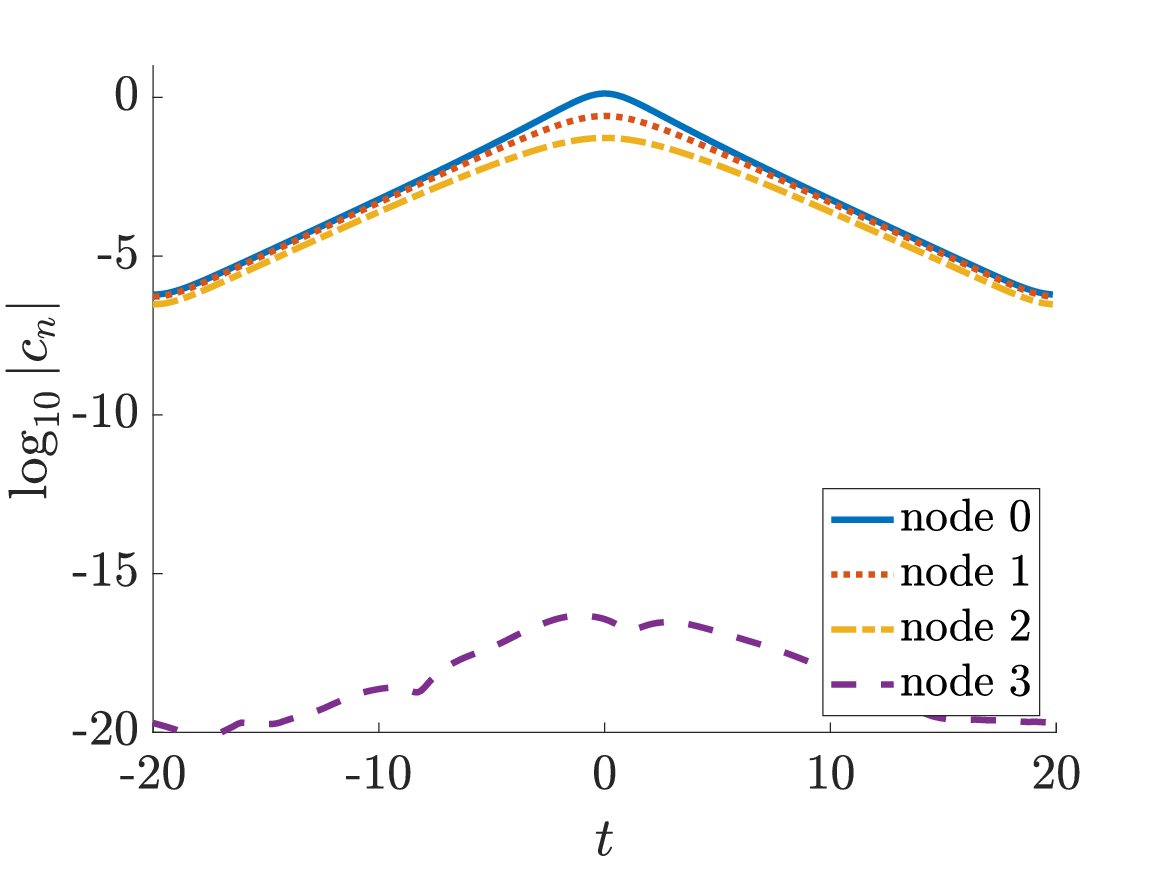}
    \end{subfigure}
        \begin{subfigure}{0.3\linewidth}
        \caption{}
        \label{fig:m6pi6phase}
        \includegraphics[width=5cm]{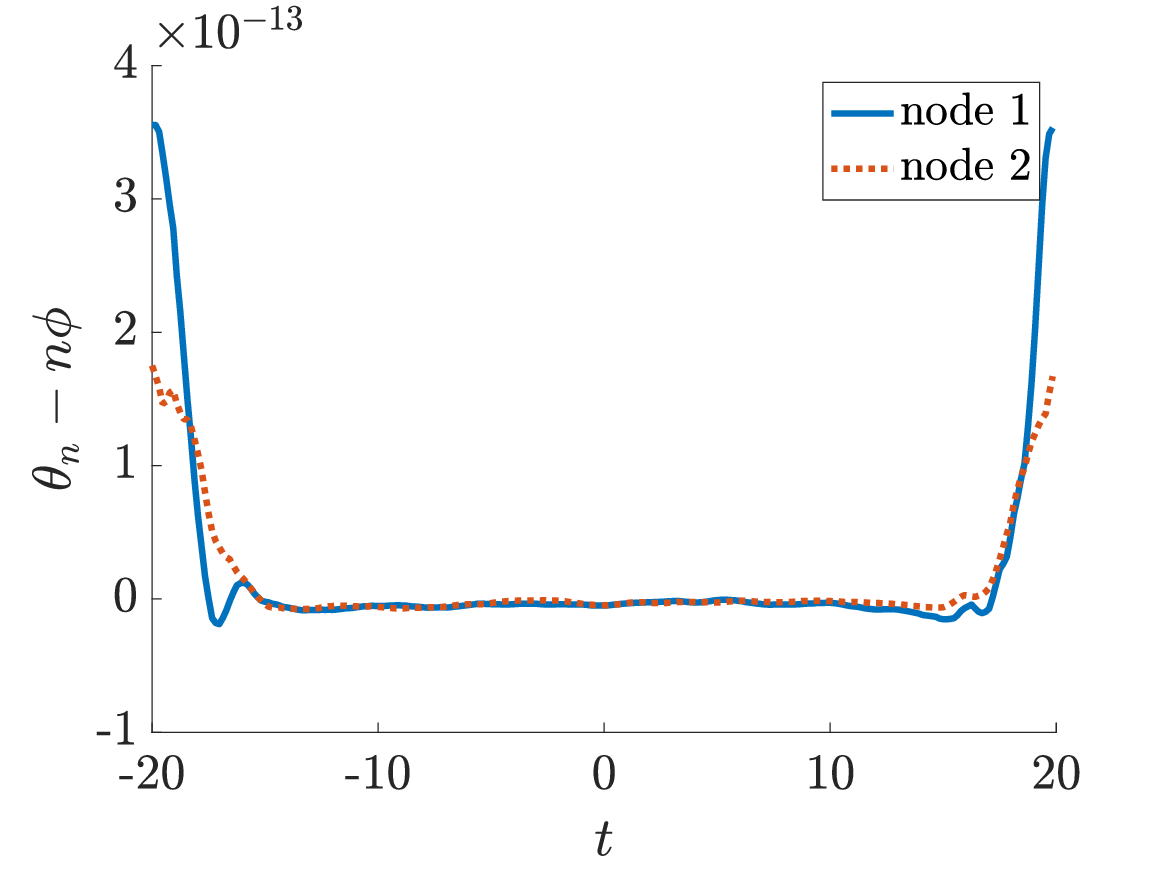}
    \end{subfigure}
    \caption{Spatiotemporal solutions to equation \cref{eq:standingwave} for $N=6$ waveguides and $k=0.25$. The twist parameter is $\phi = 0.25$ (top) and $\phi = \pi/6$ (bottom). Left and middle: Amplitude and log amplitude of solution at first four sites. Right: $\theta_n(t) - n \phi$ for sites 1 and 2.}
    \label{fig:m6sol}
\end{figure}

Now that we have constructed solutions of \cref{eq:standingwave}, we show that we can attain suppression of the opposite core when $\phi=\pi/N$. In \cref{fig:m6sol}, we show the amplitudes, $a_n$, and phases, $\theta_n$, for $N=6$ cores and $k=0.25$, for $\phi=0.25$ (top row) and $\phi=\pi/6$ (bottom row). As predicted, when $\phi=\pi/6$, there is significant suppression of the amplitude at the opposite core (compare \cref{fig:m6025logamp} and \cref{fig:m6pi6logamp}). In addition, we see that the angle deviation $\theta_n - n \phi$ is also significantly suppressed when $\phi=\pi/N$ (compare \cref{fig:m6025phase} and \cref{fig:m6pi6phase}). Although we only showed that the lowest order terms in the asymptotic expansion \cref{eq:asympsol} are suppressed when $\phi=\pi/N$, numerical evidence suggests that,
as in the model without temporal dispersion \cite{parker2021}, this suppression is complete, i.e. the amplitude of node 3 in \cref{fig:m6pi6logamp} lies at the margin of machine precision.    
In \cref{fig:m6supp}, we show how both the amplitude, $a_3$, of the opposite core and the deviations, $\phi_n - n \phi$, of the phases are suppressed as $\phi$ varies; the cusps at $\phi=\pi/6$ provides further numerical evidence for complete suppression. 
\cref{fig:suppN} shows that this suppression at $\phi=\pi/N$ persists for larger values of $N$. In particular, there is a dropoff of approximately 15 orders of magnitude between the amplitude $a_{N/2-1}$ and the amplitude $a_{N/2}$ of the opposite core. For all four values of $N$, the amplitude of the opposite core lies at the margin of machine precision.
These results do not depend on the number of Fourier nodes in the temporal discretization.

% figure: amplitudes/phases for N=6

% figure: amplitudes/phases for N=6

\begin{figure}
    \centering
    \begin{subfigure}{0.4\linewidth}
        \caption{}
        \label{fig:m6errora}
        \includegraphics[width=5cm]{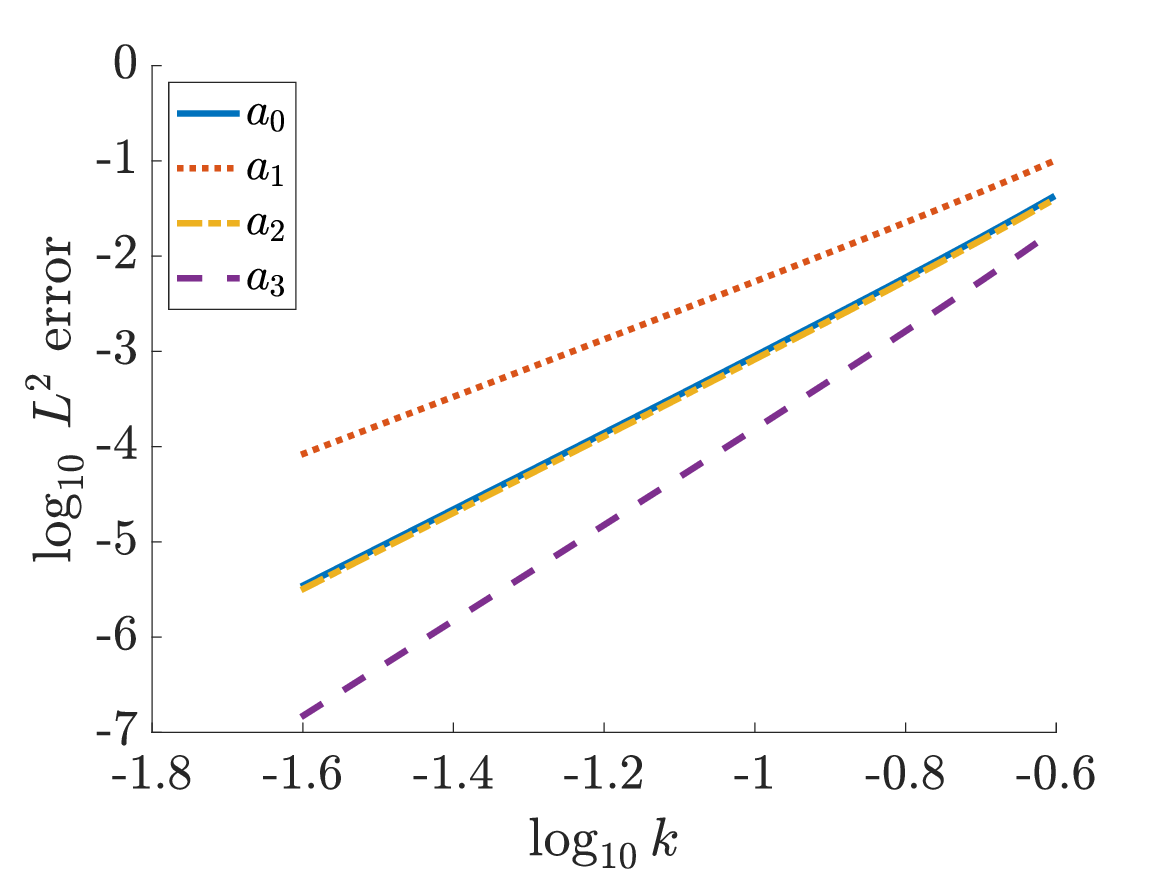}
    \end{subfigure}
    \begin{subfigure}{0.4\linewidth}
        \caption{}
        \label{fig:m6errorb}
        \includegraphics[width=5cm]{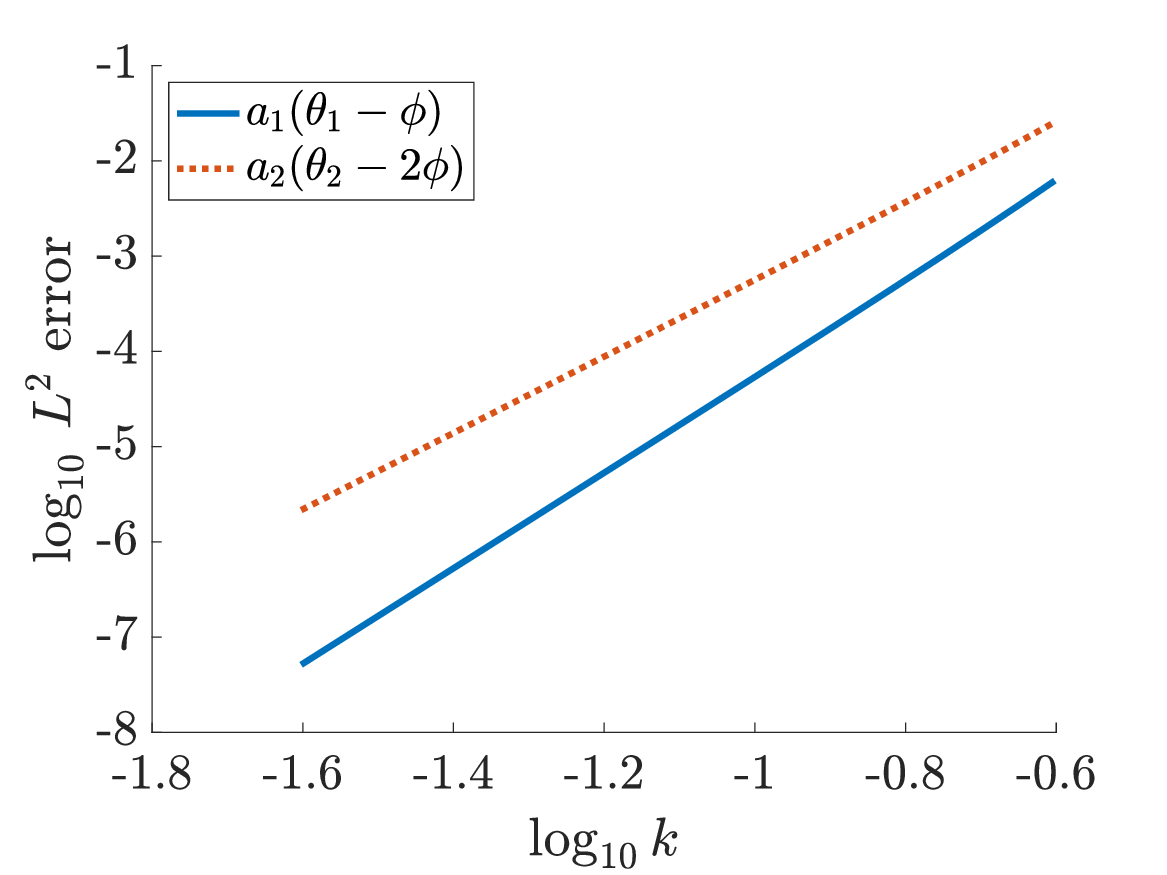}
    \end{subfigure}
    \caption{(a) $\log_{10}$ of $L^2$ error of the amplitudes $a_n$ vs $\log_{10} k$, and (b) the product of the amplitudes and phases $a_n \theta_n $ vs $\log_{10} k$. (We note that $0.25 \approx 10^{-0.6}$).}
    \label{fig:m6error}
\end{figure}

Finally, we compute the error in the asymptotic formulae for the amplitudes, $a_n$, and phases, $\theta_n$. In \cref{fig:m6errora}, we plots the log of the $L^2$ error $\| a_n^{\text{num}} - a_n^{\text{asymp}} \|_{L^2}$ vs. the log of coupling parameter $k$, where $a_n^{\text{num}}$ is obtained from numerical parameter continuation, $a_n^{\text{asymp}}$ is computed using \cref{eq:asympsol}, \cref{eq:tildeanpsi}, and \cref{eq:tildea0eq}, and the $L^2$ norm on the periodic domain $[-T,T]$ is defined by
\[
\| f \|_{L^2} = \int_{-T}^T |f(t)|^2 dt.
\] 
The slopes of least squares regressions lines for the error in $a_0$, $a_1$, $a_2$, and $a_3$ are within 5\% of 4, 3, 4, and 5 (respectively). 
For $n \geq 1$, this agrees with the order of the remainder term for $a_n$ in \cref{eq:asympsol}.
The order of the remainder term for $a_0$ is one order higher in $k$ than in \cref{eq:asympsol}. 
For $k \leq 0.2$, the relative $L^2$ error 
\[
\frac{ \| a_n^{\text{num}} - a_n^{\text{asymp}} \|_{L^2} } { | a_n^{\text{num}} \|_{L^2} }
\]
is less than 0.05.

Since the formulas \cref{eq:tildeantnpsi} involve the products, $a_n \theta_n$, of the amplitudes and phases, \cref{fig:m6errorb} plots the log of the $L^2$ error of $\| (a_n (\theta_n - n \phi)^{\text{num}} - a_n(\theta_n - n \phi)^{\text{asymp}}\|_{L^2}$ vs. the log of coupling parameter $k$, where the numerical value is obtained from numerical parameter continuation, and the asymptotic value is computed using 
$a_n (\theta_n - n \phi )^{\text{asymp}} = k^{N-n} \widetilde{a}_n \widetilde{\theta}_n$ and \cref{eq:tildeantnpsi}. The slopes of least squares regressions lines for the error in $a_1\theta_1$ and $a_2\theta_2$ are within 2.5\% of 5 and 4 (respectively), which suggests that the order of the remainder term in $k$ is correct in the asymptotic expansions \cref{eq:asympsol} for $\theta_n$.

% figure: amplitudes/phases for N=6

\section{Stability}\label{sec:stability}

To analyze the stability of a solution to \cref{eq:standingwavematrix}, we express each $c_n$ as $c_n = u_n + i v_n$, and let $c = (u_1, \dots, u_n,v_1, \dots, v_n) \in \R^{2N}$. The linearization of the PDE \cref{eq:cz} about the pulse, $c$, is the linear operator $\calL(c)$ on $H^2(\R, \R^{2N}) \subset L^2(\R, \R^{2N})$, defined by 
\begin{equation}\label{eq:linc}
\calL(c) = \begin{pmatrix}
k A_s(\phi) + 2 \diag(u_n v_n) & \partial_t^2 - \omega + k A_c(\phi) + \diag(u_n^2) + 3 \diag(v_n^2) \\
-\partial_t^2 + \omega - k A_c(\phi) - 3\diag(u_n^2) - \diag(v_n^2) &
k A_s(\phi) - 2 \diag(u_n v_n)
\end{pmatrix},
\end{equation}
where $A_c(\phi)$ and $A_s(\phi)$ are defined in \eqref{eq:realA} and \eqref{eq:imagA}. The spectrum of $\calL(c) $ is the union of the essential spectrum and the point spectrum. Since the 
pulse is even in the spatial variable, the
 essential spectrum is the set of $\lambda \in \C$ for which $\calL(c) - \lambda \calI$ is not Fredholm \cite[Section 3.1]{Kapitula2013}. The point spectrum is the set of $\lambda \in \C$ for which $\ker( \calL(c) - \lambda \calI)$ is nontrivial. 
 \begin{comment}%JZ
 (Although the essential spectrum also comprises those $\lambda \in \C$ for which $\calL(c) - \lambda \calI$ is Fredholm with nonzero index, we will explain below that this is not possible.) 
 \end{comment}
 The essential spectrum depends only on the background state, i.e. is independent of $c$. Since the operator $\calL(c)$ is exponentially asymptotic to the linear operator $\calL(0)$, given by
\begin{equation}\label{eq:L0}
\calL(0) = \begin{pmatrix}
k A_s(\phi) & \partial_t^2 - \omega + k A_c(\phi)  \\
-\partial_t^2 + \omega - k A_c(\phi) &
k A_s(\phi) 
\end{pmatrix},
\end{equation}
it follows from the Weyl essential spectrum theorem \cite[Theorem 2.2.6]{Kapitula2013} that the essential spectrum of $\calL(c)$ is given by the essential spectrum of the constant coefficient operator $\calL(0)$, which we compute in \cref{app:prop1proof}. 
\begin{comment}%JZ
Since $\calL(c)$ is exponentially asymptotic to $\calL(0)$ at both $\pm \infty$, i.e. the Morse indices of the asymptotic operator are the same at $\pm \infty$, $\calL(0) - \lambda \calI$ cannot have nonzero Fredholm index (see \cite[Lemma 3.1.10]{Kapitula2013}). 
\end{comment}
Consequently, we obtain the following result.

\begin{proposition}\label{prop:ess}
For $0 \leq \phi \leq \frac{2\pi}{N}$, define $\alpha$ in terms of $\phi$ by 
\begin{equation}\label{eq:essalpha}
\alpha = \begin{cases}
\omega - 2 k \cos\left(\phi\right), & 0 \leq \phi \leq \frac{\pi}{N}, \\
\omega - 2 k \cos\left(\frac{2\pi}{N}-\phi\right), & \frac{\pi}{N} \leq \phi \leq \frac{2\pi}{N},
\end{cases}
\end{equation}
and extend $\alpha$ periodically in $\phi$ with period $2 \pi/N$. If $\omega$, $k$, and $\phi$ are chosen so that $\alpha > 0$, the essential spectrum comprises two symmetric, disjoint bands on the imaginary axis
\begin{align}\label{eq:lessinterval}
\sigma_{\text{ess}} = i (-\infty, -\alpha] \cup i [\alpha, \infty),
\end{align}
and $\alpha$ attains a maximum of $\omega - 2 k \cos\left(\pi/N\right)$ when $\phi = \pi/N$.
\end{proposition}
We note that the essential spectrum is bounded away from the origin when $\alpha>0$. In addition, both $\alpha$ and $k^*$ attain a maximum when $\phi = \pi/N$. 

\begin{figure}
    \centering
    \begin{subfigure}{0.4\linewidth}
        \caption{}
        \label{fig:phi0speca}
        \includegraphics[width=5cm]{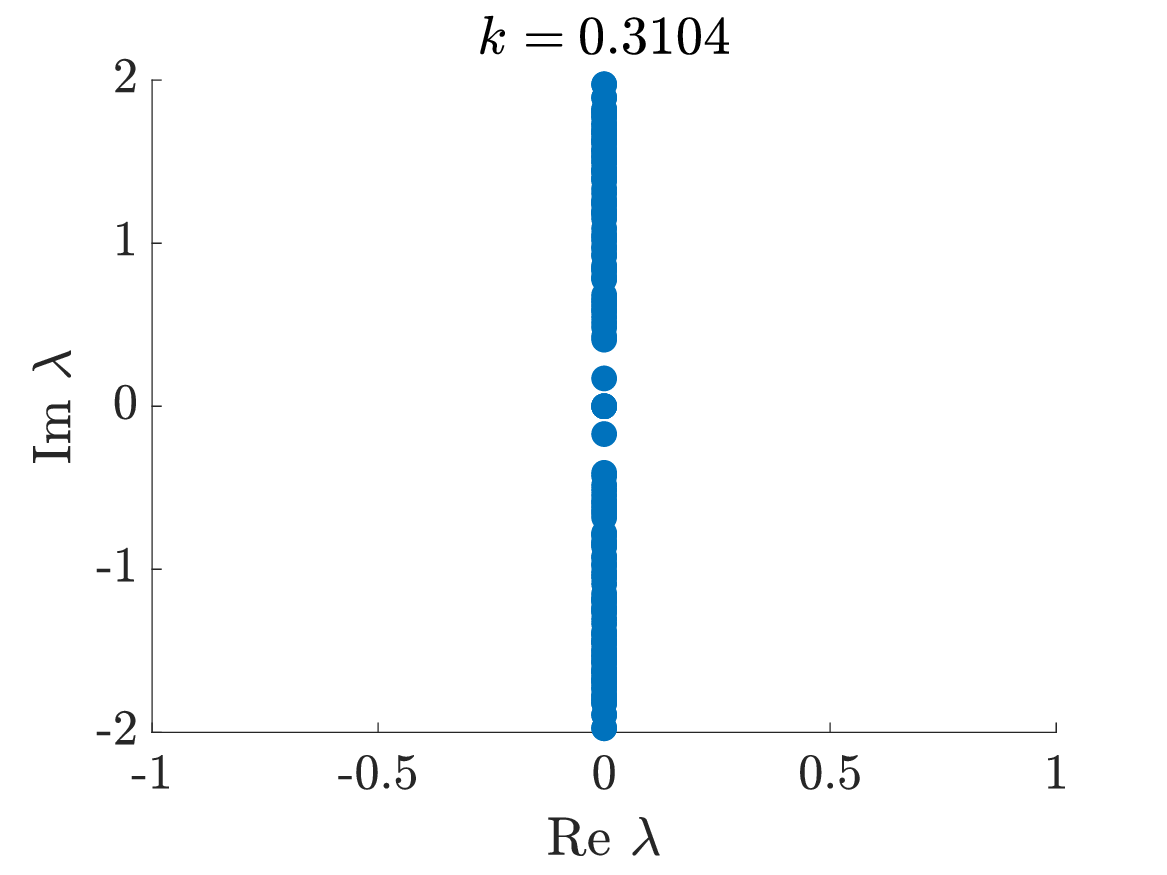}
    \end{subfigure}
    \begin{subfigure}{0.4\linewidth}
        \caption{}
        \label{fig:phi0specb}
        \includegraphics[width=5cm]{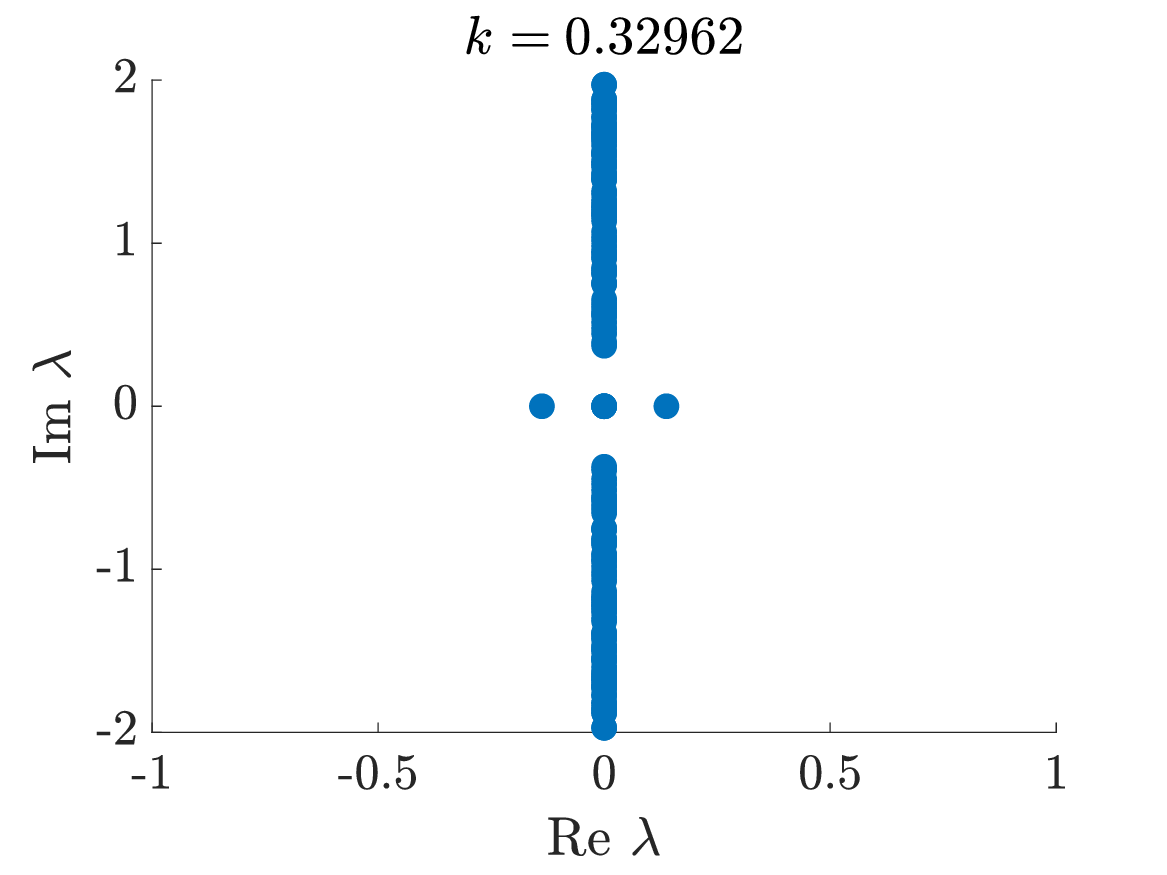}
    \end{subfigure}
    \begin{subfigure}{0.4\linewidth}
        \caption{}
        \label{fig:phi0specc}
        \includegraphics[width=5cm]{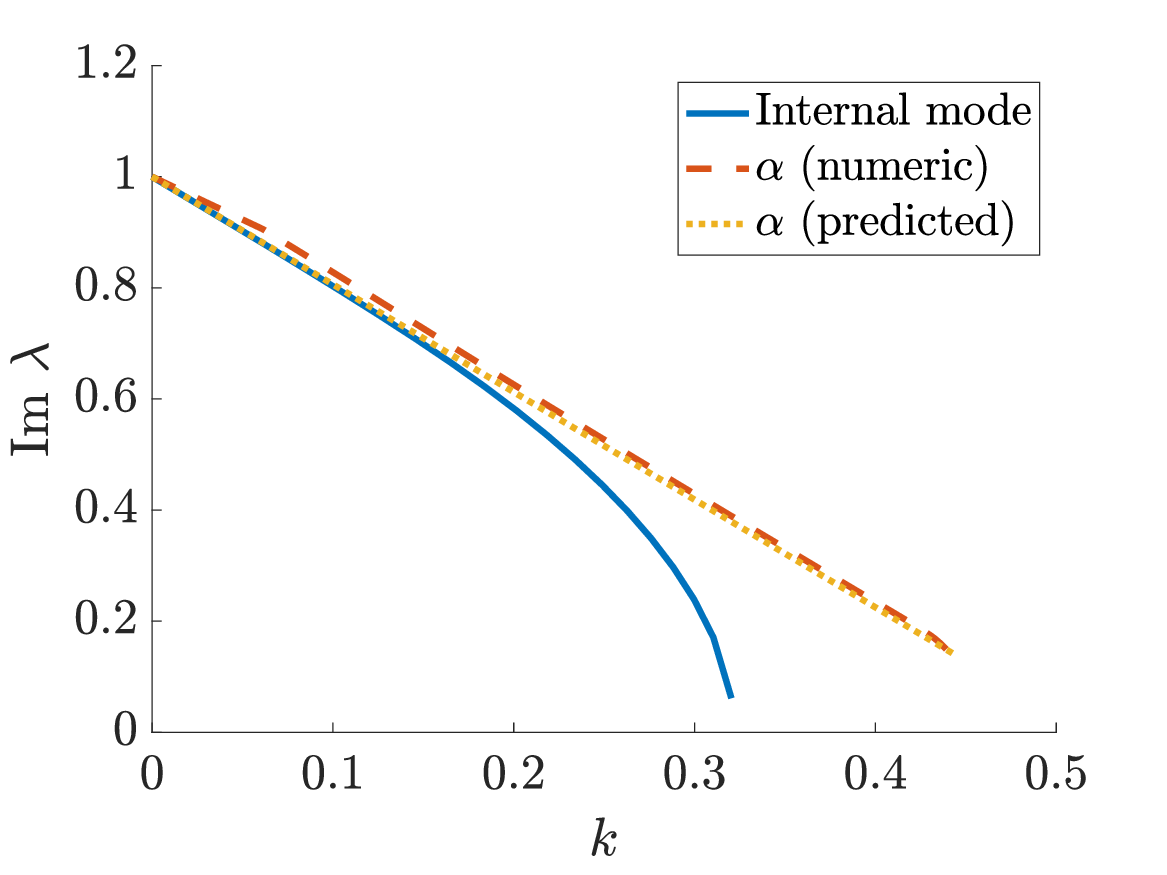}
    \end{subfigure}
    \begin{subfigure}{0.4\linewidth}
        \caption{}
        \label{fig:phi0specd}
        \includegraphics[width=5cm]{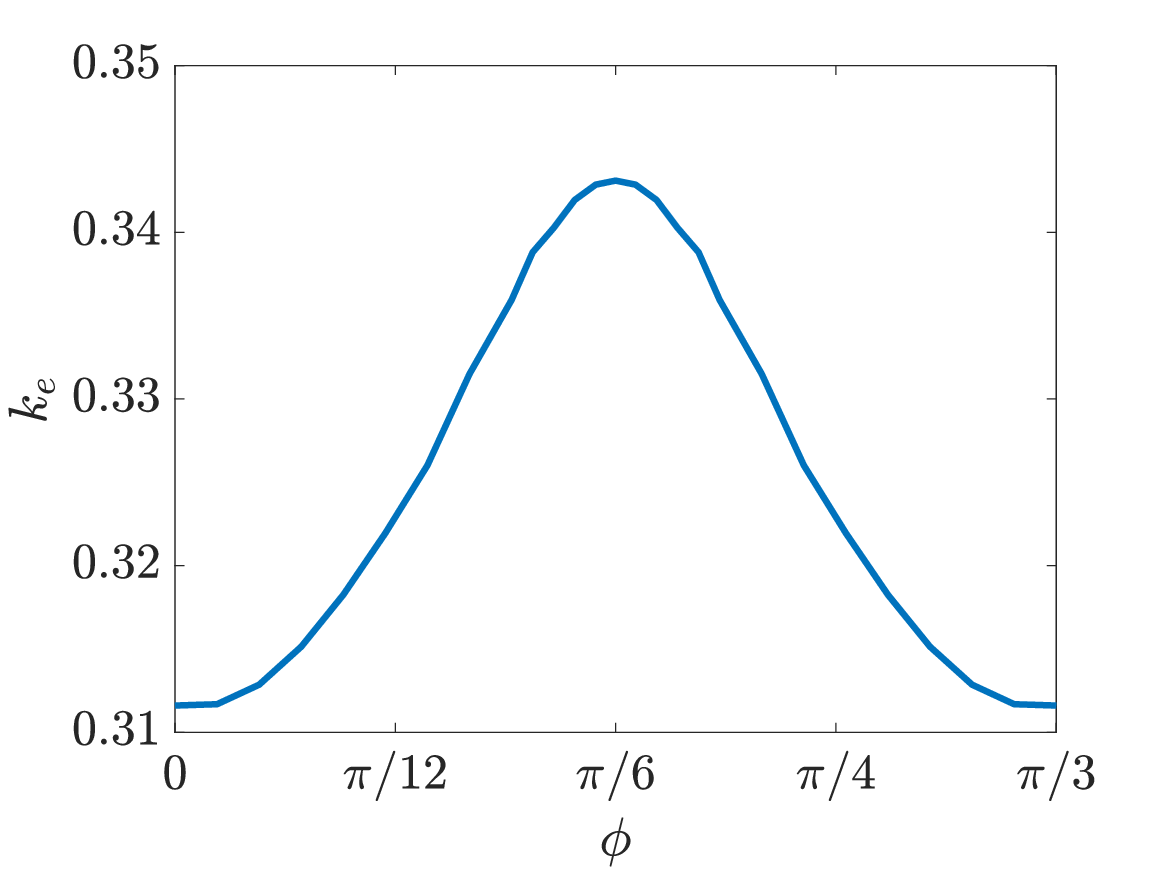}
    \end{subfigure}
    \caption{(a), (b) Spectrum of $\calL(c)$ before and after collision of internal mode eigenvalue with origin. (c) Plot of predicted essential spectrum border $\alpha$, computed essential spectrum border, and internal mode eigenvalue vs. $k$. (d) Plot of critical value $k_e$ for internal mode eigenvalue collision vs. $\phi$. The twist parameter is $\phi = 0.25$ for all plots.}
    \label{fig:phi0spec}
\end{figure}

The spectrum of $\calL(c)$ can be numerically approximated by discretizing the differentiation matrices in \cref{eq:linc} using Fourier spectral differentiation matrices, writing the operator as a block matrix, and computing the eigenvalues of that matrix using MATLAB's eigenvalue solver \texttt{eig}. 
The two essential spectrum bands can be seen in the top row of \cref{fig:phi0spec}.
Due to the spatial discretization, the continuous bands constituting the essential spectrum are approximated by discrete sets of points. In \cref{fig:phi0specc}, we plot the border, $\alpha$, of the essential spectrum as a function of the coupling parameter, $k$. The results obtained using the matrix discretization of the operator are shown with the red dashed curve, and those obtained using \cref{eq:essalpha} are shown with the yellow dotted curve. The relative error of the numerical result is less than 1\%. 

We now turn to the point spectrum of $\calL(c)$. As for the NLS equation~\cite[Chapter 2.1.1.1]{Kevrekidis2009},
the kernel of $\calL(c)$ has algebraic multiplicity 4 and geometric multiplicity 2. The kernel is spanned by the two eigenfunctions 
$w_1 = (\partial_t u_1, \dots, \partial_t u_N, \partial_t v_1, \dots, \partial_t v_N )^T$ and 
$w_2 = (-v_1, \dots, -v_N, u_1, \dots, u_N )^T$. In addition, there is a pair of internal mode eigenvalues $\pm \lambda_e$, which will call edge modes since they split off from the edge of the essential spectrum (these can be seen in \cref{fig:phi0speca}). As $k$ increases, $\lambda_e$ moves towards the origin (see \cref{fig:phi0specc} for a plot of $\lambda_e$ vs. $k$), collides with the eigenvalues at the origin at a critical value $k = k_e$, and then moves onto the real axis (\cref{fig:phi0specb}). This suggests that the solution $c$ is neutrally stable for $0 < k < k_e$ and unstable for $k > k_e$. A plot of this critical value $k_e$ vs. $\phi$ in \cref{fig:phi0specd} suggests that the maximum value of $k_e$ occurs when $\phi = \pi/N$.

\begin{figure}
    \centering
    \begin{subfigure}{0.4\linewidth}
        \caption{}
        \label{fig:evolz1a}
        \includegraphics[width=5cm]{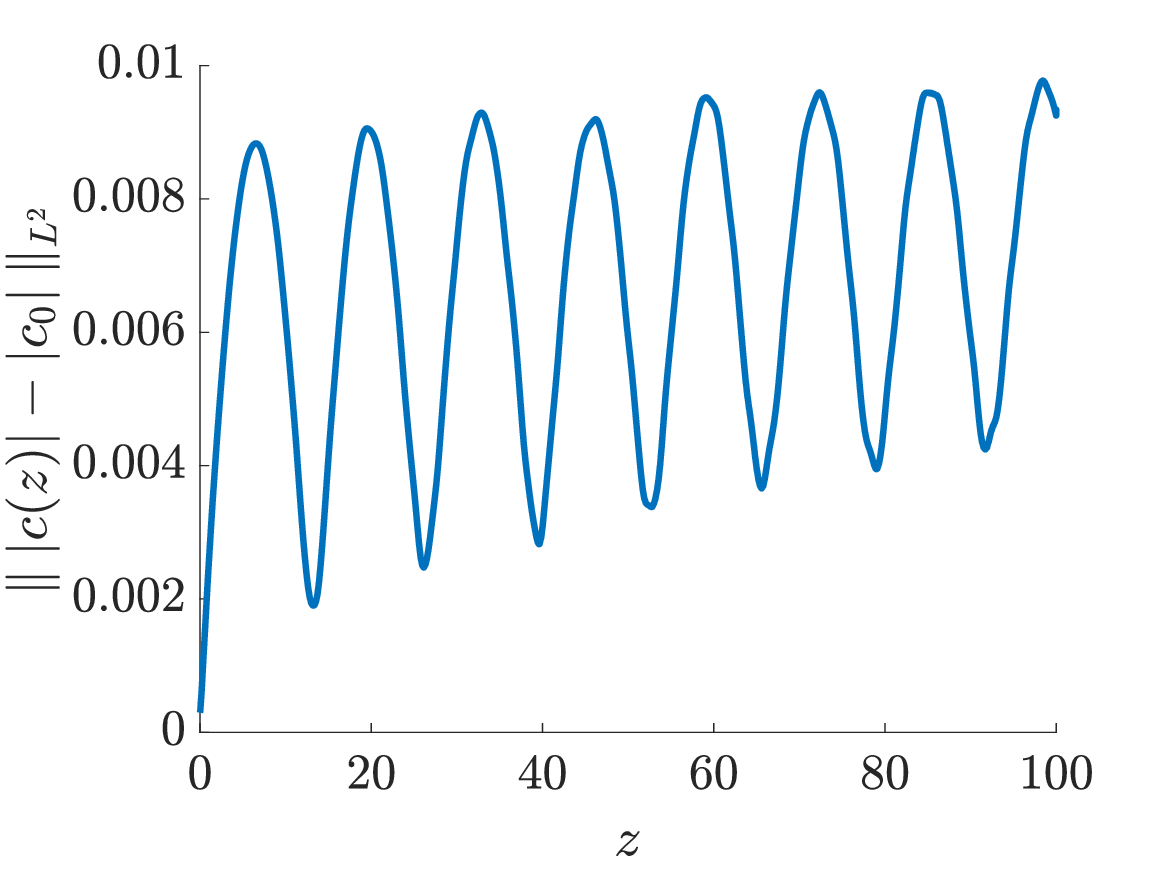}
    \end{subfigure}
    \begin{subfigure}{0.4\linewidth}
        \caption{}
        \label{fig:evolz1b}
        \includegraphics[width=5cm]{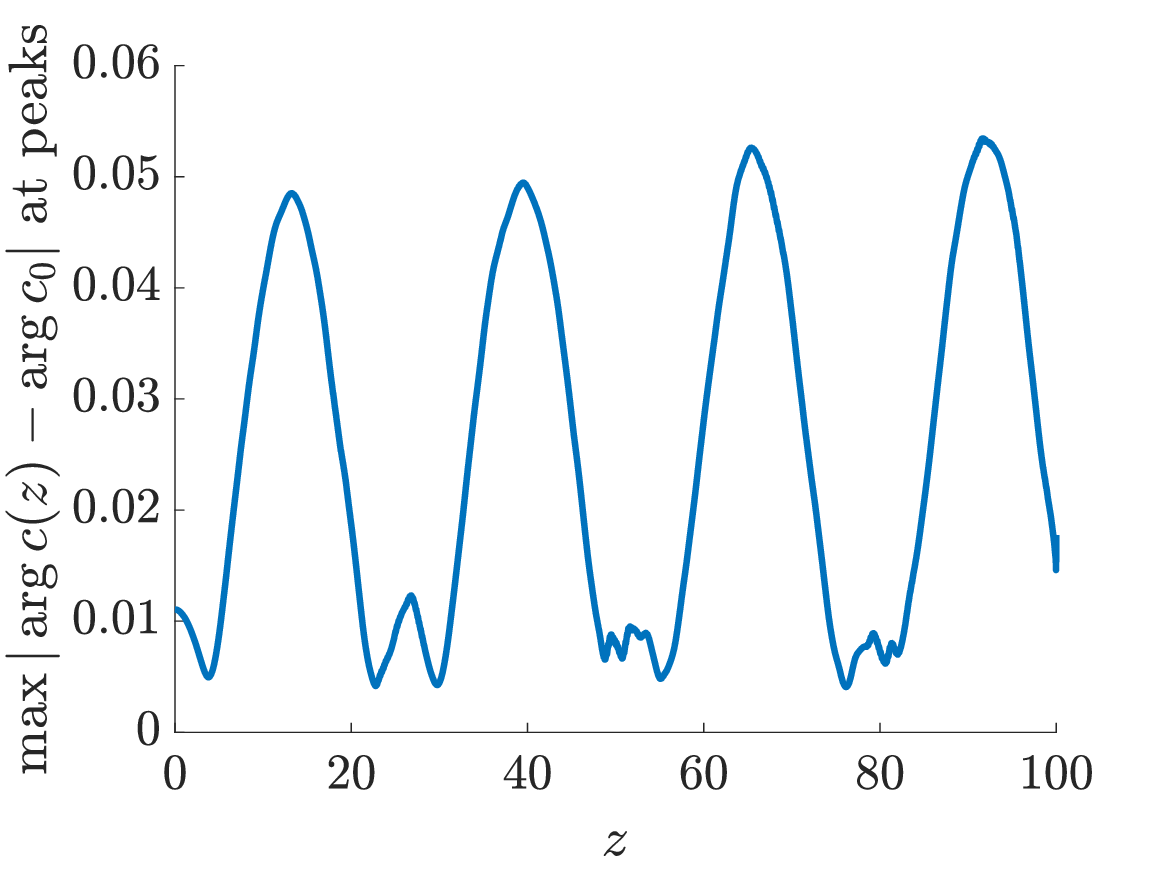}
    \end{subfigure}
        \begin{subfigure}{0.4\linewidth}
        \caption{}
        \label{fig:evolz1c}
        \includegraphics[width=5cm]{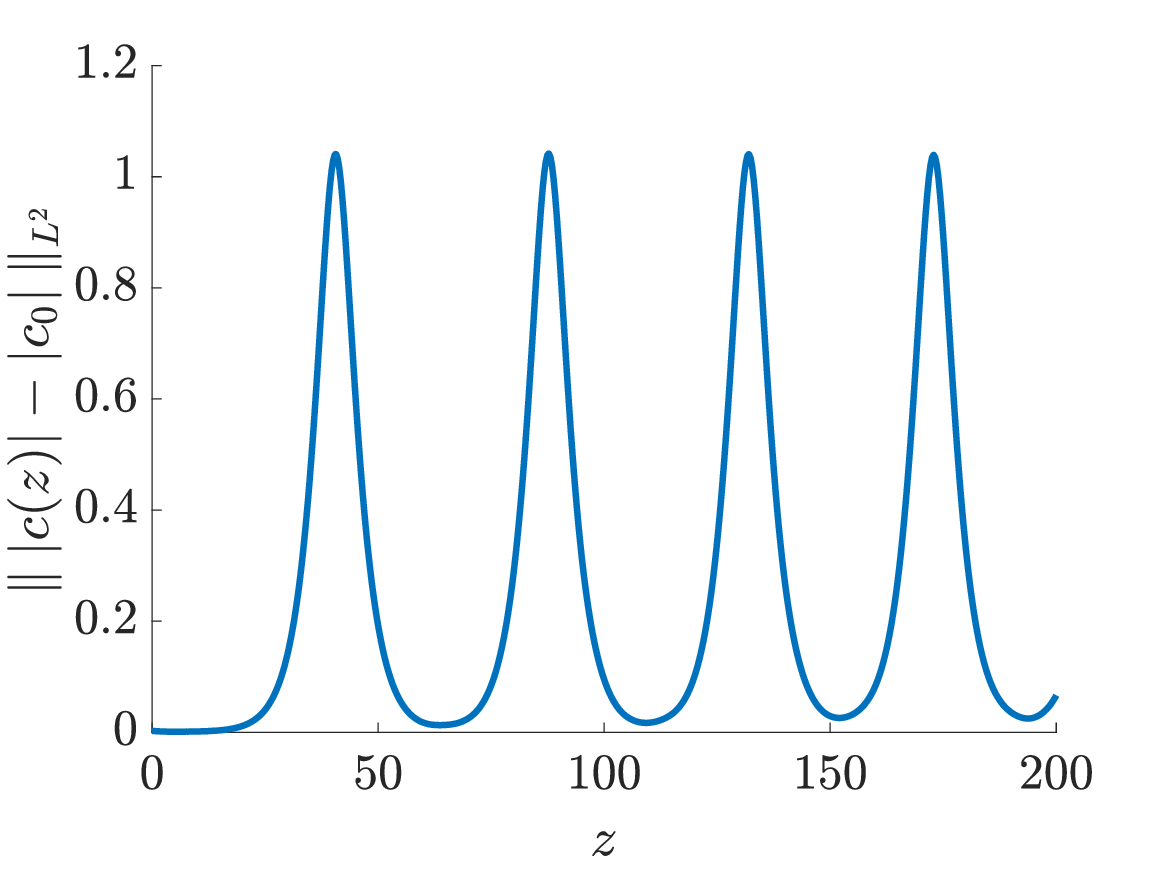}
    \end{subfigure}
    \begin{subfigure}{0.4\linewidth}
        \caption{}
        \label{fig:evolz1d}
        \includegraphics[width=5cm]{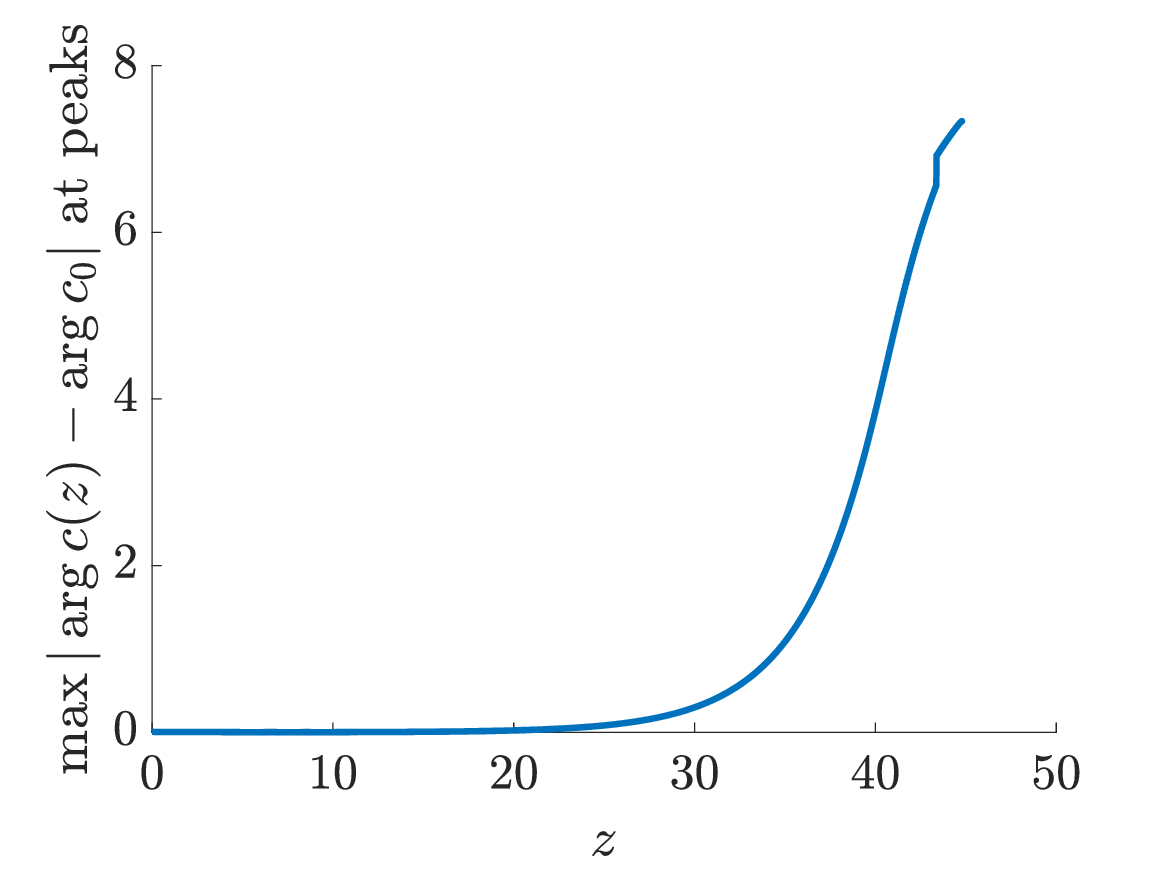}
    \end{subfigure}
    \caption{Evolution, $c(z)$, in $z$ of perturbations of the pulse solution, $c_0$, for $k = 0.25$ (top, edge mode $\lambda = \pm 0.239i$) and $k = 0.35$ (bottom, edge mode $\lambda = \pm 0.255$). The initial condition is $c_0 + \epsilon v_e$, where $v_e$ is the eigenfunction corresponding to the edge mode and $\epsilon = 10^{-2}$. Left panels: Evolution of the $L^2$ norm difference between the amplitude $|c(z)|$ of perturbation and the amplitude $|c_0|$ of the pulse. Right panels: Evolution of phase differences at centers of peaks between the perturbation $c(z)$ and the pulse $c_0$. Twist parameter $\phi = 0.25$ for all plots.}
    \label{fig:evolz1}
\end{figure}

Numerical evolution experiments using MATLAB's \texttt{ode45} function suggest that this internal mode primarily affects the phases, $\theta_n$, of the pulses. For $k < k_e$, the internal mode eigenvalue, $\lambda_e$, is imaginary. Evolution of perturbations in the direction of the corresponding eigenmode $v_e$ exhibits phase oscillations with frequency, $\lambda_e$, with relative error less than $10^{-3}$ (see \cref{fig:evolz1b}). For $k > k_e$, the internal mode eigenvalue, $\lambda_e$, is real. Evolution of perturbations in the direction of the corresponding eigenmode, $v_e$, exhibits exponential growth in the phases; the growth rate is $\lambda_e$ with relative error less than $10^{-2}$ (see \cref{fig:evolz1d}). We note that in both cases, the amplitudes of the perturbed solutions also exhibit oscillatory behavior (\cref{fig:evolz1a} and \cref{fig:evolz1c}), although the magnitude of these oscillations in $L^2$ norm is much larger for $k > k_e$ than for $k < k_e$.

\section{Conclusions}

In this paper, we studied a model of light propagation in a twisted multi-core fiber that includes temporal dispersion. We derived asymptotic expressions for spatiotemporal solutions in which the number of waveguides, $N$, is even, the bulk of the intensity is contained in a single node (the primary core), and the amplitudes and phases of the solution obey certain symmetry relations. 
This asymptotic analysis is valid for arbitrary, even $N$.
These expressions are in very good agreement with results from numerical parameter continuation. Furthermore, the asymptotic analysis shows that for an even number, $N$, of waveguides, there is significant suppression of optical activity in the waveguide opposite the primary core when the twist parameter, $\phi$, and the number of waveguides, $N$, are related by $\phi = \pi/N$. This is the exact same behavior found in the model with no temporal dispersion \cite{parker2021}.
Numerical computation suggests that there are two critical values, $k_e$ and $k^*$, of $k$, with $0 < k_e < k^*$. The pulse solution exists for $0 < k < k^*$ and is stable for $0 < k < k_e$, at which point an internal eigenmode collides with the eigenvalues at the origin. Both $k_e$ and $k^*$ attain their maximum when $\phi = \pi/N$, which is also the value of the twist parameter for which the essential spectrum boundary, $\alpha$, is furthest from the origin.

Future research could include exploring the case with odd $N$, which was considered in the model without temporal dispersion. As in \cite{parker2021}, this would involve a different set of symmetries, and it may be possible that a configuration could be found in which there would be optical suppression of a single site when $\phi = \pi/N$.
Furthermore, we only considered solutions which obey the symmetries \cref{eq:symm}. Solution branches with these symmetries cease to exist for $k > k^*$, where $k^*$ is the point at which all cores have attained equal amplitudes. It is possible that there exist asymmetric solutions for which $k > k^*$.
We could also study multi-pulses, which are solutions in which the energy is concentrated at multiple sites in the ring. These have been well-studied in DNLS \cite{Parker2020}, and we considered them briefly in \cite{parker2021}. Finally, if the distances between neighboring waveguides are small, nonlocal interactions might become significant, suggesting a model with coupling that extends beyond nearest neighbors. We could, for example, incorporate next-to-next-nearest neighbor interactions, as was done with DNLS in \cite{penati2018}.

It could also be interesting to extend the model to more complicated geometries, such as twisted optical fibers comprising multiple concentric rings, or to study the evolution of these solutions in the presence of imperfections in the fiber. Another avenue of exploration would be to examine the continuum limit, i.e. what happens as $N \rightarrow \infty$. Since the fiber radius $R$ is held constant, the ring becomes more ``densely'' packed with waveguides as $N$ increases, and in the limit as $N \rightarrow \infty$, this becomes a continuum on the circle of radius $R$. Furthermore, the twist parameter, $\phi$, is proportional to $1/N$, and so $\phi \rightarrow 0$ as $N \rightarrow \infty$. An interesting question is to what extent the twist is preserved in the equations for the continuum limit. Finally, since these solutions are stable for $k < k_e$, it would be interesting compare these results with experimental realizations.

\vspace{0.5cm}
% \begin{acknowledgments}
\textbf{Acknowledgments}: 
This material is based upon work supported by the U.S. National Science Foundation under the RTG grant DMS-1840260 (R.P. and A.A.), DMS-1909559 (A.A.), and DMS-2106203 (J.Z.). 
% \end{acknowledgments}

\appendix

\section{Symmetries}\label{app:symm}

We verify that the symmetry relations in \cref{eq:symm} are consistent with \cref{eq:st2}, i.e. solutions to \cref{eq:st2} can exist with these symmetries. First, for $n = 2, \dots, N/2-1$, we take equation \cref{eq:st2} for $-n$, substitute the symmetries for $a_n$ and $\theta_n$ from \cref{eq:symm}, and simplify, to obtain 
\begin{equation*}
\begin{aligned}
(\ddot a_n &- a_n (\dot \theta_n)^2) 
- i ( a_n \ddot\theta_n + 2 \dot a_n \dot \theta_n )\\
&+ k\left(a_{n-1}e^{i[(\theta_n - \theta_{n-1}) - \phi]} + a_{n+1}e^{-i[(\theta_{n+1} - \theta_{n}) - \phi]} \right)+a_n^3 - \omega a_n = 0,
\end{aligned}
\end{equation*}	
which is the complex conjugate of \cref{eq:st2} for $n$. For $n = 0$, we take equation \cref{eq:st2} for $n=0$, substitute the symmetries for $a_n$ and $\theta_n$ from \cref{eq:symm}, and simplify, to obtain 
\begin{equation*}
\begin{aligned}
(\ddot a_0 &- a_0 (\dot \theta_0)^2) 
+ i ( a_0 \ddot\theta_0 + 2 \dot a_0 \dot \theta_0 )
+ 2 k a_1 \cos(\theta_1 - \phi)(\cos \theta_0 + i \sin \theta_0) + a_n^3 - \omega a_n = 0.
\end{aligned}
\end{equation*}
The imaginary part is
\begin{equation}\label{eq:n0imagpart}
a_0 \ddot\theta_0 + 2 \dot a_0 \dot \theta_0 = 2 k a_1 \cos(\theta_1 - \phi) \sin \theta_0,
\end{equation}
for which $\theta_0 = 0$ is a solution, thus this condition is consistent. Following the same procedure by using the imaginary part of equation \cref{eq:st2}, we can show that the $\theta_{N/2} = 0$ is a solution to the imaginary part of \cref{eq:st2} for $n=N/2$, thus this condition is consistent as well.

\section{Asymptotic analysis}\label{app:asymp}

Let $a_n$ and $\theta_n$ be the amplitudes and phases at site $n$, as in the ansatz \cref{eq:cnansatz}, where $n \in S = \{ 0, \pm 1, \dots, \pm N/2-1, N/2 \}$. We assume that 
\[
a_n, \theta_n \in H^2(\R) \subset L^2(\R),
\]
where $L^2(\R)$ is the space of real, valued square-integrable functions on $\R$ equipped with the standard inner product, and $H^2(\R)$ is the corresponding Sobolev space. We note that the system of equations \cref{eq:st2} is translation invariant, i.e. if $\{ a_n(t), \theta_n(t)\}_{n\in S}$ is a solution, then so is $\{ a_n(t-\tau), \theta_n(t-\tau)\}_{n\in S}$ for any $\tau \in \R$. When $k = 0$, we want the solution $a_0$ to be the ordinary NLS soliton $\psi$, which is centered at 0. To ensure that the solution we find is not shifted in $t$, we impose the phase condition,
\begin{equation}\label{eq:phasecond}
\langle a_0, \dot{\psi} \rangle_{L^2(\R)} = \int_{-\infty}^\infty a_0(t) \dot{\psi}(t) dt = 0,
\end{equation}
on the amplitude $a_0$. In other words, $a_0$ has no component in the direction of $\dot{\psi}$.

We define the linear operator $\Lw: H^2(\R) \subset L^2(\R) \rightarrow L^2(\R)$ by
\begin{equation}\label{eq:Lw}
\Lw = \omega - \partial_t^2.
\end{equation}
The kernel of $\Lw$ is $\{0\}$, i.e. the only solution in $H^2(\R)$ to $\Lw f = 0$ is $f = 0$, which can be verified by taking the Fourier transform of $\Lw$. Consequently, the operator $\Lw$ is invertible on $L^2(\R)$. We also note that if we use periodic boundary conditions on $[-T,T]$, as long as we take $\omega > 0$, $\ker \Lw = \{0\}$, and $\Lw$ is invertible on $L^2_{\text{per}}([-T,T])$.

Next, we recall that the NLS soliton, $\psi$, is a real-valued, standing wave solution to the NLS equation with frequency $\omega$, thus it solves 
\begin{equation}\label{eq:NLSreal}
\ddot{u} + u^3 - \omega u = 0. 
\end{equation}
Linearizing this equation about $\psi$ yields the self-adjoint linear operator $\calL(\psi): H^2(\R) \subset L^2(\R) \rightarrow L^2(\R)$, defined by
\begin{equation}\label{eq:Lpsi}
\calL(\psi) = \partial_t^2 - \omega + 3 \psi^2.
\end{equation}
The kernel of $\calL(\psi)$ is one-dimensional, and is spanned by $\dot{\psi}$. By the Fredholm alternative \cite{Ramm2001}, since $\calL(\psi)$ is self-adjoint, the equation $\calL(\psi) = u$ has a solution if $u \perp \dot{\psi}$.

\subsection{Primary core}

Using the symmetry relations \cref{eq:symm} and $\theta_0 = 0$, equation \cref{eq:st2real} for $n=0$ becomes
\begin{equation}\label{eq:st2reala0}
\ddot a_0  + 2 k a_1 \cos(\theta_1 - \phi) + a_0^3 - \omega a_0 = 0.
\end{equation}
We take the power series ansatz 
\[
a_0 = \psi + k a_0^{(1)} + k^2 a_0^{(2)} + \mathcal{O}(k^3)
\]
for the amplitude $a_0$, where $\psi$ is the NLS soliton \cref{eq:NLSsoliton}. In addition, from \cref{eq:basicseries}, we have, at minimum, 
\[
a_1 = k \widetilde{a}_1 + \mathcal{O}(k^2), \qquad \theta_1 = \phi + \mathcal{O}(k).
\]
Substituting this into \cref{eq:st2reala0}, using the Taylor series expansion $\cos(\theta_1-\phi) = \mathcal{O}(k^2)$, collecting powers of $k$, and simplifying, we obtain
\begin{equation*}
\begin{aligned}
&\left(\ddot{\psi} + \psi^3 - \omega \psi\right) 
+ k\left(\ddot a_0^{(1)} - \omega a_0^{(1)} + 3 \psi^2 a_0^{(1)}\right) \\
&\qquad\qquad+ k^2\left(\ddot a_0^{(2)} - \omega a_0^{(2)} + 3\left(a_0^{(1)}\right)^2 + 3 \psi^2 a_0^{(2)} + 2 \widetilde{a}_1 \right) + \mathcal{O}(k^3) = 0.
\end{aligned}
\end{equation*}
The $\mathcal{O}(1)$ term is 0 since $\psi$ solves \cref{eq:NLSreal}. The $\mathcal{O}(k)$ term can be written as $\calL(\psi)a_0^{(1)}=0$, where $\calL(\psi)$ is defined by \cref{eq:Lpsi}. Since the kernel of $\calL(\psi)$ is spanned by $\dot \psi$, $a_0^{(1)} = c \dot \psi$ for some constant $c$. However, since we wish our solution to satisfy the phase condition \cref{eq:phasecond}, we require that $a_0^{(1)} = 0$. The $\mathcal{O}(k^2)$ term then becomes
\[
\ddot a_0^{(2)} - \omega a_0^{(2)} + 3 \psi^2 a_0^{(2)} + 2 \widetilde{a}_1 = 0.
\]
Letting $\widetilde{a}_0 = a_0^{(2)}$, we rewrite this as
\begin{align}\label{eq:solvea0ta}
\calL(\psi) \widetilde{a}_0 = -2 \widetilde{a}_1,
\end{align}
which is the first equality in \cref{eq:tildea0eq}. Once we determine $\widetilde{a}_1$, which we will do in the next step, we can solve for $\widetilde{a}_0$, provided $\widetilde{a}_1 \perp \dot\psi$. Putting all of this together, we have the expression for $a_0$
\begin{equation}\label{eq:a0eq}
a_0 = \psi + k^2 \widetilde{a}_0 + \mathcal{O}(k^3) = \psi + \mathcal{O}(k^2),
\end{equation}
which is the first equation in \cref{eq:asympsol}.

\subsection{Amplitudes}\label{sec:amplitudes}

We can now determine the leading order terms for the amplitudes, $a_n$, for $n = 1, \dots, N/2$, using the real parts \cref{eq:st2real} of the equations \cref{eq:st2}. The phases $\theta_n$ will be determined in the next step. For $n=1$, we take the power series ansatz 
\[
a_1 = k \widetilde{a}_1 + \mathcal{O}(k^2).
\]
From \cref{eq:basicseries}, we have, at minimum,
\[
a_2 = \mathcal{O}(k^2), \qquad \theta_1 = \phi + \mathcal{O}(k), \qquad \theta_2 = 2 \phi + \mathcal{O}(k),
\]
and we note that $\dot \theta_1 = \mathcal{O}(k)$. Substituting these together with the expression \cref{eq:a0eq} for $a_0$ into the $n=1$ equation of \cref{eq:st2real}, expanding the cosine terms in a Taylor series, collecting powers of $k$, and simplifying, we obtain the equation
\[
k\left(\partial_t^2 \widetilde{a}_1 - \omega \widetilde{a}_1 + \psi\right) + \mathcal{O}(k^2) = 0.
\]
We note that the nonlinear term $a_1^3$ does not contribute to the lowest order term in the asymptotic expansion. We use the $\mathcal{O}(k)$ term to solve for $\widetilde{a}_1$ to obtain
\begin{equation}\label{eq:a11}
\widetilde{a}_1 = (\omega - \partial_t^2)^{-1} \psi,
\end{equation}
which has a solution since $\Lw$ is invertible. This is the first equation in \cref{eq:tildean}. Therefore,
\begin{equation}\label{eq:a1eq}
a_1 = k (\omega - \partial_t^2) \psi + \mathcal{O}(k^2).
\end{equation}

We continue this process iteratively from $n=2$ to $n=N/2-1$. We take the power series ansatz 
\[
a_n = k^n \widetilde{a}_n + \mathcal{O}(k^{n+1}),
\]
and use
\begin{align*}
&a_{n-1} = k^{n-1} \widetilde{a}_{n-1} + \mathcal{O}(k^{n}), &&a_{n+1} = \mathcal{O}(k^{n+1}), \\
&\theta_{n-1} = (n-1) \phi + \mathcal{O}(k), &&\theta_{n+1} = (n+1) \phi + \mathcal{O}(k),
\end{align*}
where the expression for $a_{n-1}$ was found in the previous step. Substituting these into \cref{eq:st2real} and following the same procedure as above, we obtain the equation
\begin{align*}
k^n\left(\partial_t^2 \widetilde{a}_n - \omega \widetilde{a}_n + \widetilde{a}_{n-1} \right) +\mathcal{O}(k^{n+1}) = 0.
\end{align*}
We use the $\mathcal{O}(k^n)$ term to solve for $\widetilde{a}_n$ to obtain
\begin{equation}\label{eq:ann}
\widetilde{a}_n = (\omega - \partial_t^2)^{-1}\widetilde{a}_{n-1},
\end{equation}
which is the second equation in \cref{eq:tildean}. This gives us
\begin{equation}\label{eq:aneq}
a_n = k^n (\omega - \partial_t^2)^{-1} \widetilde{a}_{n-1} + \mathcal{O}(k^{n+1}).
\end{equation}
We will show in \cref{sec:hot} that the remainder term is $\mathcal{O}(k^{n+2})$.

For the opposite core, $n=N/2$, we take $\theta_{N/2} = 0$ in the real part of \cref{eq:st2} for $n = N/2$, use the symmetry relations for $a_{N/2-1}$ and $\theta_{N/2-1}$ from \cref{eq:symm}, and simplify to obtain
\begin{equation}\label{eq:oppcore}
\ddot a_{N/2} - a_{N/2} (\dot \theta_{N/2})^2 + 
2 k a_{N/2-1}\cos( \theta_{N/2-1} + \phi) + a_{N/2}^3 - \omega a_{N/2} = 0.
\end{equation}
Using the power series ansatz
\[
a_{N/2} = k^{N/2} \widetilde{a}_{N/2} + \mathcal{O}(k^{N/2+1})
\]
and the formula $a_{N/2-1} = k^{N/2-1} \widetilde{a}_{N/2-1} + \mathcal{O}(k^{N/2})$ from the previous step, we obtain the expression
\begin{equation}\label{eq:am2a}
\widetilde{a}_{N/2} = 2 (\omega - \partial_t^2)^{-1}\cos( \theta_{N/2-1} + \phi) \widetilde{a}_{N/2-1}.
\end{equation}
It is important to note that since $\theta_{N/2} = 0$, the cosine term in \cref{eq:am2a} is order $\mathcal{O}(1)$. Substituting the ansatz $\theta_{N/2-1} = (N/2-1)\phi + \mathcal{O}(k)$ and simplifying, equation \cref{eq:am2a} becomes
\begin{equation}\label{eq:am2}
\widetilde{a}_{N/2} = 2 \cos( N\phi/2)(\omega - \partial_t^2)^{-1} \widetilde{a}_{N/2-1} + \mathcal{O}(k),
\end{equation}
which is the third equation in \cref{eq:tildean}. Since $\widetilde{a}_{N/2} = \mathcal{O}(1)$, it follows that
\begin{equation}\label{eq:am2eq}
a_{N/2} = 2 k^{N/2} \cos\left( N\phi / 2\right) (\omega - \partial_t^2)^{-1} \widetilde{a}_{n-1} + \mathcal{O}(k^{N/2+1}).
\end{equation}
Finally, now that we have derived an expression \cref{eq:a11} for $\widetilde{a}_1$, we substitute this into \cref{eq:solvea0ta} to obtain 
\[
\calL(\psi) \widetilde{a}_0 = -2 (\omega - \partial_t^2)^{-1} \psi,
\]
which we can solve for $\widetilde{a}_0$. This is the second equality in \cref{eq:tildea0eq}.

\subsection{Phases}

Now that we have determined the leading order terms for the amplitudes, $a_n$, we will compute the leading order terms for the phases, $\theta_n$. In fact, we will get expressions for the products $\widetilde{a}_n \widetilde{\theta}_n$ of the leading order terms of the amplitudes and the phases. To do this, we will use the imaginary parts \cref{eq:st2imag} of the equations \cref{eq:st2}, this time working backwards from $n = N/2-1$ to $n=1$. (Equations \cref{eq:st2} have already been satisfied for $n=N/2$ and $n=0$ by taking $\theta_{N/2} = 0$ and $\theta_0 = 0$, respectively). For $n=N/2-1$, substitute the power series ansatz
\[
\theta_{N/2-1} = k^2 \widetilde{\theta}_{N/2-1} + \mathcal{O}(k^{3}),
\]
the expressions for $a_n$ from the previous section, $\theta_{N/2} = 0$, and
$\theta_{N/2-2} = (N/2-2)\phi + \mathcal{O}(k^4)$ from \cref{eq:basicseries} into equation \cref{eq:st2} for $n=N/2-1$ to obtain
\[
k^{N/2+1} \left( \widetilde{a}_{N/2-1} \ddot{\widetilde{\theta}}_{N/2-1} + 2 \dot{\widetilde{a}}_{N/2-1} \dot{\widetilde{\theta}}_{N/2-1} - \widetilde{a}_{N/2-2} \widetilde{\theta}_{N/2-1} - \widetilde{a}_{N/2} \sin(N \phi/2) \right) + \mathcal{O}(k^{N/2+2}) = 0.
\]
We use the $\mathcal{O}(k^{N/2+1})$ term to solve for $\widetilde{\theta}_{N/2-1}$. First, we solve equation \cref{eq:ann} for $\widetilde{a}_{N/2-2}$ and substitute this above to obtain
\[
\widetilde{a}_{N/2-1} \ddot{\widetilde{\theta}}_{N/2-1} + 2 \dot{\widetilde{a}}_{N/2-1} \dot{\widetilde{\theta}}_{N/2-1} - \widetilde{\theta}_{N/2-1} (\omega - \partial_t^2) \widetilde{a}_{N/2-1} - \widetilde{a}_{N/2} \sin(N \phi/2) = 0,
\]
which simplifies as the derivative of a product to obtain
\[
(\omega - \partial_t^2)\left( \widetilde{a}_{N/2-1} \widetilde{\theta}_{N/2-1} \right) = -\sin(N \phi/2) \widetilde{a}_{N/2}.
\]
Applying $(\omega - \partial_t^2)^{-1}$ to both sides, we obtain the formula
\begin{equation}\label{am2thm2}
\widetilde{a}_{N/2-1} \widetilde{\theta}_{N/2-1} = -\sin(N \phi/2) (\omega - \partial_t^2)^{-1} \widetilde{a}_{N/2},
\end{equation}
and we can divide by $\widetilde{a}_{N/2-1}$ to solve for $\widetilde{\theta}_{N/2-1}$. Using equation \cref{eq:am2} for $\widetilde{a}_{N/2}$, equation \cref{am2thm2} becomes
\[
\widetilde{a}_{N/2-1} \widetilde{\theta}_{N/2-1} = -2 \sin(N \phi/2) \cos( N\phi/2) (\omega - \partial_t^2)^{-2} \widetilde{a}_{N/2-1},
\]
which simplifies to 
\begin{equation}\label{am2thm2a}
\widetilde{a}_{N/2-1} \widetilde{\theta}_{N/2-1} = -\sin(N \phi) (\omega - \partial_t^2)^{-2} \widetilde{a}_{N/2-1},
\end{equation}
using the double angle formula. This is the first equation in \cref{eq:tildeantn}.

We now follow an iterative procedure from $n=N/2-2$ down to $n=1$. For $n$, substitute the power series ansatz
\[
\theta_{n} = k^{N-2n} \widetilde{\theta}_{n} + \mathcal{O}(k^{N-2n+1}),
\]
the expressions for $a_n$ from the previous section, and $\theta_n$ from \cref{eq:basicseries} into equation \cref{eq:st2imag} for $n$, and simplify, to obtain
\[
k^{N/2} \left( \widetilde{a}_n \ddot{\widetilde{\theta}}_n + 2 \dot{\widetilde{a}}_n \dot{\widetilde{\theta}}_n
- \widetilde{a}_{n-1} \widetilde{\theta}_n + \widetilde{a}_{n+1}\widetilde{\theta}_{n+1} \right) + \mathcal{O}\left(k^{N/2+1}\right) = 0.
\]
We use the $\mathcal{O}(k^{N/2})$ term to solve for $\widetilde{\theta}_{n}$. First, we solve equation \cref{eq:ann} for $\widetilde{a}_{n-1}$ and substitute this above to obtain
\[
\widetilde{a}_n \ddot{\widetilde{\theta}}_n + 2 \dot{\widetilde{a}}_n \dot{\widetilde{\theta}}_n
- \widetilde{\theta}_n (\omega - \partial_t^2) \widetilde{a}_n + \widetilde{a}_{n+1} \widetilde{\theta}_{n+1} = 0,
\]
which simplifies as the derivative of a product to obtain
\[
(\omega - \partial_t^2)\left( \widetilde{a}_n \widetilde{\theta}_n \right) = \widetilde{a}_{n+1} \widetilde{\theta}_{n+1}.
\]
Applying $(\omega - \partial_t^2)^{-1}$ to both sides, we obtain the formula
\begin{equation}\label{an2th}
\widetilde{a}_n \widetilde{\theta}_n = (\omega - \partial_t^2)^{-1} \left( \widetilde{a}_{n+1} \widetilde{\theta}_{n+1} \right),
\end{equation}
which is the second equation in \cref{eq:tildeantn}. We can divide by $\widetilde{a}_n$ to solve for $\widetilde{\theta}_n$. 

\subsection{Higher order terms}\label{sec:hot}

We briefly consider the higher order terms in the asymptotic expansion for the amplitudes, $a_n$. This will give us estimates for the order of the remainder term in the asymptotic expansion for $a_n$ which are slightly better than we obtained in \cref{sec:amplitudes}. We also obtain an improved estimate for the remainder term in the opposite core when $\phi=\pi/N$.
We take the asymptotic series ansatz
\begin{align}\label{eq:seriesansatz}
a_n &= k^n \sum_{j=0}^\infty a_n^{(j)} k^j && n = 0, \dots, \frac{N}{2},
\end{align}
where, using the notation in the previous section for the leading order terms, we have $a_0^{0}= \psi$, $a_0^{1}=0$, $a_0^{2} = \widetilde{a}_0$, and $a_n^{0} = \widetilde{a}_n$ for $n = 1, \dots, N/2$. Substituting this into \cref{eq:st2real}, using the estimate in \cref{eq:basicseries} for $\theta_n$, and simplifying, we obtain the following expression for the next-to-leading order terms:
\begin{align*}
\calL(\psi) a_0^{(3)} &= -2 a_1^{(1)} \\
a_n^{(1)} &= (\omega - \partial_t^2)^{-1}a_{n-1}^{(1)} && n = 1, \dots, N/2-1 \\
a_{N/2}^{(1)} &= 2 \cos( N\phi/2)(\omega - \partial_t^2)^{-1} a_{N/2-1}^{(1)}.
\end{align*}
Since $a_0^{(1)}=0$, $a_n^{(1)}=0$ for $n = 1, \dots, N/2$, from which it follows that  
$a_n = k^n \widetilde{a}_n + \mathcal{O}(k^{n+2})$ for $n \geq 1$, which is the second equation in \cref{eq:asympsol}.
In principle, we can continue this process to obtain still higher order correction terms, although we note that the expressions we obtain will become increasingly more unwieldy. The next term in the asymptotic series, for example, is
\begin{align*}
a_n^{(2)} &= (\omega - \partial_t^2)^{-1}(a_{n-1}^{(2)} + a_{n+1}^{0}) && n = 1, \dots, N/2-1,
\end{align*}
which involves a contribution from both neighboring nodes. Higher terms in the series will incorporate contributions from the nonlinearity as well as the phase $\theta_n$.

For the opposite core, we substitute the power series ansatz \cref{eq:seriesansatz}, $\theta_{N/2}=0$, and the estimate $\theta_{N/2-1} = \mathcal{O}(k^2)$ from \cref{eq:tildeantn} into \cref{eq:oppcore}, expand the cosine term in a Taylor series, and simplify to obtain
\[
\sum_{j=0}^\infty a_{N/2}^{(j)} k^j + 2 \left[\cos( N\phi/2) + \mathcal{O}(k^4) \right] \sum_{j=0}^\infty a_{N/2-1}^{(j)} k^j - \omega \sum_{j=0}^\infty a_{N/2}^{(j)} + \mathcal{O}(k^N) = 0.
\]
Since $N$ is even, and we will always take $N\geq 4$, the $\mathcal{O}(k^4)$ terms will only contribute for $j \geq 4$, thus we obtain
\begin{align*}
a_{N/2}^{(j)} &= 2 \cos( N\phi/2)(\omega - \partial_t^2)^{-1} a_{N/2-1}^{(j)} && j = 0,1,2,3.
\end{align*}
It follows that if $\phi=\pi/N$, then $a_{N/2}^{(j)}=0$ for $j = 0,1,2,3$, in which case we obtain the estimate
\[
a_{N/2} = \mathcal{O}\left({k^{N/2+4}}\right).
\]

\section{Proof of \texorpdfstring{\cref{prop:ess}}{Proposition 1}}\label{app:prop1proof}

Following \cite[Section 3.1]{Kapitula2013} (see, in particular, Section 3.1.1.5, concerning the NLS equation), $\lambda \in \sigma_{\text{ess}}(\calL(0))$ if and only if $\calL(0) - \lambda$ has an imaginary spatial eigenvalue, i.e. 
$\left(\calL(0)-\lambda \calI \right) \begin{pmatrix}u\\v\end{pmatrix} e^{i r t}$
has a solution for $r \in \R$ and $u, v \in \C^N$. After simplification, this reduces to the system of equations 
\begin{equation}\label{eq:L0system}
\begin{aligned}
&(k A_s - \lambda)u - (r^2 + \omega - k A_c) v = 0 \\
&(r^2 + \omega - k A_c)u + (k A_s - \lambda) v = 0.
\end{aligned}
\end{equation}
$A_c$ and $A_s$ are both circulant matrices \cite{davis2012circulant}. The eigenvalues, $\mu_j^c$ and $\mu_j^s$, of the circulant matrices $A_c$ and $A_s$, are given by
\begin{equation}\label{eq:circeigs}
\begin{aligned}
\mu_j^c &= 2 \cos\left( \frac{2 \pi}{N} j\right) \cos \phi, \qquad 
\mu_j^s = 2 i \sin\left( \frac{2 \pi}{N} j\right) \sin \phi, \qquad
j = 0, \dots, N-1,
\end{aligned}
\end{equation}
and the corresponding common eigenvectors are given by
\begin{equation}\label{eq:circevecs}
v_j = \frac{1}{\sqrt{N}}\left(1, \xi^j, \xi^{2j}, \dots, \xi^{(N-1)j} \right),
\end{equation}
where $\xi = \exp{\frac{2\pi i}{N}}$ is a primitive $N$th root of unity \cite{davis2012circulant}. Since $A_s$ and $A_c$ appear in both equations in \cref{eq:L0system}, we can only have a solution if $u$ and $v$ are scalar multiples of one of the eigenvectors $v_j$. Letting $u = b_1 v_j$ and $v = b_2 v_j$, the system of equations \cref{eq:L0system} reduces to 
\begin{equation}\label{eq:L0system2}
\begin{aligned}
&(k \mu_j^s - \lambda) b_1 - (r^2 + \omega - k \mu_j^c) b_2 = 0 \\
&(r^2 + \omega - k \mu_j^c) b_1 + (k \mu_j^s - \lambda) b_2 = 0,
\end{aligned}
\end{equation}
where we used the fact that $v_j \neq 0$. This has a nontrivial solution if and only if the determinant of the corresponding $2 \times 2$ coefficient matrix is 0, i.e.
\begin{equation*}
(\lambda - k \mu_j^s)^2  + (r^2 + \omega - k \mu_j^c)^2 = 0.
\end{equation*}
For $k \mu_j^c \leq \omega$, this has a solution when
\begin{align*}
\lambda &= k \mu_j^s  \pm i (r^2 + \omega - k \mu_j^c) \\
&= \pm i \left( r^2 + \omega \pm 2 k \sin\left( \frac{2 \pi}{N} j\right) \sin \phi 
- 2 k \cos\left( \frac{2 \pi}{N} j\right) \cos \phi
\right) \\
&= \pm i \left( r^2 + \omega - 2k \cos \left( \frac{2 \pi}{N} j \pm \phi \right) \right).
\end{align*}
As long as $2k \max_{j=0,\dots,N-1} \cos \left( \frac{2 \pi}{N} j \pm \phi \right) \leq \omega$, $\lambda \in (-\infty,-\alpha]\cup[\alpha, \infty)$, where
\begin{align}\label{eq:alpha1}
\alpha = \omega - 2k \max_{j=0,\dots,N-1} \cos \left( \frac{2 \pi}{N} \right).
\end{align}
For $0 \leq \phi \leq 2 \pi/N$, 
\begin{align*}
\alpha = \begin{cases}
\omega - 2 k \cos\left(\phi\right) & 0 \leq \phi \leq \frac{\pi}{N} \\
\omega - 2 k \cos\left(\frac{2\pi}{N}-\phi\right) & \frac{\pi}{N} \leq \phi \leq \frac{2\pi}{N},
\end{cases}
\end{align*}
which has a maximum of $\omega - 2 k \cos\left(\pi/N\right)$ when $\phi = \pi/N$. For $\phi$ outside of this interval, $\alpha$ is periodic with period $\frac{2\pi}{N}$. 

\bibliographystyle{amsplain}
\bibliography{twist2.bib}

\end{document}